\documentclass{aa}  

\usepackage{graphicx}
\usepackage{txfonts}

\usepackage{natbib}
\usepackage{natbib,twoopt}
\usepackage[breaklinks=true]{hyperref} %% to avoid \citeads line fills
\bibpunct{(}{)}{;}{a}{}{,}
\usepackage{color}

\begin{document}

   \title{SUDARE-VOICE variability-selection of active galaxies in the \emph{Chandra} Deep Field South and the SERVS/SWIRE region}
  %% shorter title for page header
   \titlerunning{Variability-selection of Active Galaxies in the CDFS area}

  % \subtitle{Variable Active Galaxies in the Chandra Deep Field South.}

   \author{S. Falocco
          \inst{1,2}
          \and
          M. Paolillo\inst{1,2,3}
          \and
          G. Covone \inst{1,2}
\and
 D. De Cicco \inst{1}
\and
G. Longo \inst{1,2}
\and
A. Grado\inst{4}
\and
L. Limatola\inst{4}
\and
M. Vaccari\inst{8,14}
\and
M.T. Botticella \inst{4}
\and
G. Pignata\inst{5,6}
\and
E. Cappellaro\inst{7}
\and
D. Trevese\inst{9}
\and
F. Vagnetti\inst{10}
\and
M. Salvato\inst{11}
\and
M. Radovich\inst{7}
\and
L. Hsu\inst{11}
\and 
M. Capaccioli\inst{1,2,4}
\and
N. Napolitano \inst{4}
\and
W. N. Brandt \inst{12,13}
\and
A. Baruffolo \inst{7}
\and
E. Cascone \inst{4}
\and
P. Schipani \inst{4}
% L. Greggio \inst{6}
          }

   \institute{Physics department, University Federico II,   Via Cintia, 80126, Naples, Italy
\and
INFN Napoli, Via Cintia, 80126, Naples, Italy
\and
Agenzia Spaziale Italiana Science Data Center, Via del Politecnico snc, I-00133
Roma, Italy
\and
INAF Osservatorio Di Capodimonte Naples, Italy
\and
Departamento de Ciencias Fisicas, Universidad Andres Bello, Santiago, Chile
\and
Millennium Institute of Astrophysics, Santiago, Chile
\and
INAF- Osservatorio di Padova, Italy
\and
Astrophysics Group, Physics Department, University of the Western Cape, Private Bag X17, 7535, Bellville, Cape Town, South Africa
\and
Department of Physics University La Sapienza Roma, Italy
\and
Department of Physics University Tor Vergata Roma, Italy
\and
Max Plank Institute fur Extraterrestrische Physik, Garching, Germany
\and
Department of Astronomy and Astrophysics, The Pennsylvania State University, University Park, PA 16802, USA
\and
Institute for Gravitation and the Cosmos, The Pennsylvania State University, University Park, PA 16802, USA
\and
INAF - Istituto di Radioastronomia, via Gobetti 101, 40129 Bologna, Italy
}

              %\email{falocco@fisica.unina.it}

   \date{Received; accepted }

% \abstract{}{}{}{}{} 
% 5 {} token are mandatory
 
  \abstract
   {One of the most peculiar characteristics of active galactic nuclei (AGNs) is their variability over all wavelengths. This property has been used in the past to select AGN samples and is foreseen to be one of the detection techniques applied in future multi-epoch surveys, complementing photometric and spectroscopic methods. }
   {In this paper, we aim to construct and characterise an AGN sample using a multi-epoch dataset in the r band from the SUDARE-VOICE survey. 
}
   {Our work makes use of the VST monitoring programme of an area surrounding the Chandra  Deep Field South to select variable sources. We use data spanning a six-month period over an area of 2 square degrees, to identify AGN based on their photometric variability. }
   { The selected sample includes 175 AGN candidates with magnitude r $<$ 23 mag. We distinguish different classes of variable sources through their lightcurves, as well as X-ray, spectroscopic, SED, optical, and IR information overlapping with our survey.}
   {We find that 12\% of the sample (21/175) is represented by Supernovae (SN). Of the remaining sources, 4\% (6/154) are stars, while 66\% (102/154) are likely AGNs based on the available diagnostics. We estimate an upper limit to the contamination of the variability selected AGN sample $\simeq$ 34\%, but we point out that restricting the analysis to the sources with available multi-wavelength ancillary information, the purity of our sample is close to 80\% (102 AGN out of 128 non-SN sources with multi-wavelength diagnostics).  Our work thus confirms the efficiency of the variability selection method, in agreement with our previous work on the COSMOS field. In addition we show that the variability approach is roughly consistent with the infrared selection.
}


   \maketitle
%
%________________________________________________________________

\section{Introduction}\label{introduction}

Active galactic nuclei (AGNs) are the most luminous persistent \footnote{Fast transients, such as supernovae (SNe) or gamma ray bursts (GRB) can emit more energy on short timescales.} sources in the Universe, and their emission is considered to be produced through accretion onto a super massive black hole (SMBH) \citep{shakura1976}. 
Traditionally AGNs have been selected and classified on the basis of their optical emission lines (e.g. \citealt{veilleux1987}). However, the discovery and early investigation of these sources was based on broad-band photometry and the characteristic UV excess that originates  in the accretion disk \citep{markarian1967}. Since then, optical colours have been broadly used to select AGNs and are expected to play a major role in future astronomical surveys, such as those foreseen for LSST (Brandt et al. 2002). As a larger portion of the electromagnetic spectrum became accessible to astronomical observations, additional selection methods have been devised.

X-ray emission, at least at bright luminosities and moderate gas column densities ($L_X>10^{42}$ erg $s^{-1}$ and $n_H<10^{23}$ $cm^{-2}$), is a most direct evidence of the presence of ongoing mass accretion. %% This has thus been used to select AGNs, including those affected by optical obscuration or Low Luminosity AGNs (LLAGNs) whose optical emission is polluted by the host galaxy light.
At low X-ray luminosities, starburst galaxies may contaminate the samples of AGN selected in soft X-rays (0.5 - 2. keV) (see e.g. \cite{cervino2002}, \cite{jimenezbailon2005}, \citealt{lehmer2012}) but such contamination is reduced above 2 keV \citep{perez1996}.
For this reason, hard X-ray surveys represent an effective method to provide a census of AGN, especially at high redshift where soft X-rays are shifted outside the observing band. For instance, deep observations of the Chandra Deep Field South (CDFS) area have been made by Chandra \citep{xue2011,giacconi,luo} and XMM-Newton \citep{comastri2010}, allowing AGNs to be identified down to $L_X=10^{41}$ erg $s^{-1}$ up to redshift $\sim$1.
%% The \cite{xue2011} survey is based on a 4 Ms observation of the CDFS: it allows to survey AGN down to 10$^{41}$ erg/s up to redshift $\sim$1.
The Swift Burst Alert Telescope has provided an all-sky survey in the hard X-rays (above 2 keV) allowing absorbed AGN (especially for z$<$0.2) and AGN to be sampled with X-ray luminosities $L_X$, spanning values from 10$^{42}$ erg/s to 10$^{44}$ erg/s for z$<$0.02 \citep{ajello2012}.

To overcome the biases introduced by dust and gas obscuration, which will affect UV, optical, and X-ray selection methods in different ways, infrared (IR) observations are commonly used, since the dusty torus that makes AGN difficult to select at ultraviolet and optical wavelengths emits radiation between 1.5 and 100 $\mu$m \citep{sanders1989}.
For instance, \cite{lacy2004} used mid-IR fluxes to construct diagnostics in order to separate AGN and galaxies in the Spitzer Space Telescope First Look Survey, relying on the different temperatures of dust in the circumnuclear and star-forming regions. The use of IR colour-selection criteria was refined in \cite{stern2005}, which reached an 80\% completeness (of the spectroscopically identified unabsorbed AGN sample) with less than 20\% contamination. 
By studying the large multi-wavelength data set in the Chandra/SWIRE Survey (0.6 $deg^2$ in the Lockman Hole), \cite{polletta2006} found a population of highly absorbed (Compton-thick) AGNs with this method. On the other hand, they found that the completeness of their IR-selected sample is 40\% with respect to the X-ray selected heavily absorbed AGN in that field, indicating that a large number of these sources are elusive even in the mid-IR.

An alternative approach for searching AGN is based on the fact that the luminosity of most AGN intrinsically vary at all wavelengths (see e.g. \cite{kawaguchi,paolillo2004,garcia-gonzalez2014,ulrich1997} and  references therein),
 thus making variability one of the most distinctive properties of these sources. It is known for causality arguments that the fluctuations observed on timescales of days and months come from relatively small regions, such as the accretion disk and/or the torus. 
Although the physical interpretation of AGN variability is still poorly understood, in radio-quiet AGN, the accretion rate and instabilities in the accretion flow seem to play a fundamental role (see, e.g. \cite{peterson2001}; and references therein); in jet-dominated sources, on the other hand,  variability can be produced by relativistic effects  (see e.g. \citealt{ulrich1997,gopal-krishna1991,qian1991,peterson2001}).

The variability selection method is thus based on the assumption that all AGN vary intrinsically in the observed band, without requiring assumptions on the spectral shape, colours, and/or line ratios. On the other hand, the method is biased against absorbed AGNs (e.g. Type II) since their primary emission from the nucleus is absorbed along the line of sight, thus any optical variability from these sources is very hard to detect. 
Variability selection helps in selecting those AGNs that can be confused with stars or galaxies of similar colours: the combination of variability selection technique with the multi-band photometry allows separating AGN from this class of contaminants \citep{ivezic2003,young2012,antonucci2014}.

Several authors \citep{hawkins1983,trevese1989,veron1995,bershady1998,geha2003,sesar2007,graham2014} have used variability-selection techniques to verify the completeness of colour-colour and spectroscopic selection techniques for high luminosity AGNs, for which the luminosities of the nuclei are well above those of the host galaxies. Other works \citep{sarajedini2003,sarajedini2006,sarajedini2011,villforth2012} extend such analysis to low-luminosity AGNs. Variability selection methods have revealed to be useful to complete the demography of low-luminosity AGNs. This is in part because variability helps to detect AGN against the host galaxy contamination and also because, as shown by \cite{trevese1994}, for example, variability is intrinsically stronger in lower luminosity sources.

In this paper we aim at constructing a new variability-selected AGN sample exploiting the data from the ongoing SUDARE-VOICE survey performed with the VLT Survey Telescope (VST). 
SUDARE (\cite{botticella2013} and Cappellaro et al., in prep.) is a VST monitoring survey aimed at searching for supernovae (SNe) at intermediate redshift, through observations in the g, r, and i bands in the region of the sky surrounding the CDFS (see Fig. 1). The VOICE survey (Vaccari et al., in prep.) is an independent effort aimed at providing UV-optical (u, g, r) coverage of two selected areas in the southern hemisphere: the extended CDFS (ECDFS) and the ELAIS South 1. The two surveys have joined efforts on the ECDFS to optimise the use of VST telescope time, providing both time-resolved and deep optical-UV observations over several square degrees to achieve different scientific goals.

A companion paper by \cite{decicco2014} (hereafter Paper I), studied the variability-selected sources in the COSMOS region (1$\times$1 $deg^2$) monitored by the SUDARE survey. 
In Paper I, the sample of variable sources was validated mainly through a comparison with X-ray-selected samples. 
The present paper focuses on the part of the SUDARE-VOICE monitoring programme covering 2 $\rm{deg}^2$ around the CDFS.
The region studied in the present paper has limited X-ray coverage, but has a significant overlap with the IR and optical surveys SWIRE \citep{lonsdale2004} and SERVS \citep{mauduit2012}, which provide excellent optical/IR ancillary data.
Since the area studied here is larger than the COSMOS region, this survey is best suited to strengthening the results of Paper I and to extending the comparison of variability selection with other optical and IR selection techniques.

\begin{figure}
\includegraphics[angle=0,width=10cm]{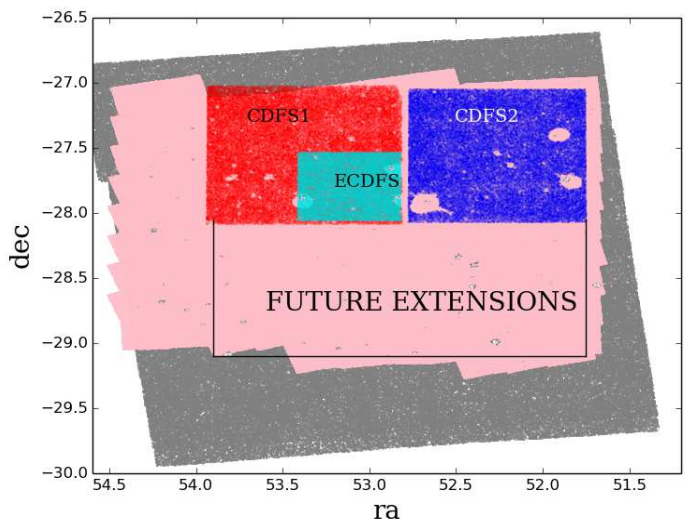}
      \caption{Region covered by the VST observations compared to other overlapping surveys. Red: VST-CDFS1, Blue: VST-CDFS2, Grey: SWIRE, Pink: SERVS, Cyan: ECDFS area. The holes in the CDFS1 and CDFS2 are due to the masks we applied to exclude bright stars, spikes, and reflection features, as explained in detail in Sect. 3. 
              }
         \label{fig:overlap}
   \end{figure}

The  paper is structured as follows. In Sect. \ref{data} we describe the
survey, in Sect. \ref{method} the methodology used for the data analysis, in Sect. \ref{results} the selection of variable sources, and in Sect. \ref{discussion} the validation of the catalogue, and
finally our conclusions and the summary are given in Sect. \ref{conclusions}.
In the following we use the AB system, unless stated otherwise.

%__________________________________________________________________

%\section{VST consortium data and public surveys in the CDFS area}\label{data}
\section{Data}\label{data}
This work is based on data in the $r$ band from the SUDARE-VOICE survey performed with the VST telescope (\cite{botticella2013}; Cappellaro et al., 2015, in prep;  Vaccari et al. 2015, in prep.). VST telescope is equipped with the OmegaCAM detector \citep{kuijken2011}, composed of 32 CCDs with a total field-of-view (FOV) of 1$^o\times1^o$.
The data are centred on the Chandra Deep Field South,
covering an area of 2 $\rm deg^2$, in the $u$, $g$, $r$, $i$ bands. 
Table \ref{tdata} summarises the observations in the $r$ band used in this work. 
The observations are spread over two adjacent fields (labelled here CDFS1 and CDFS2), and as the table shows, they span five months in the CDFS1 and four for CDFS2. The observations at each epoch are composed of five dithers with an exposure time of 30 minutes in total. Hereafter, we refer to each epoch with the corresponding VST observing block number (OB). 
The observations in the $r$ band were performed approximately every three days, avoiding the ten days around the full moon. Observations in the g and i filters were taken every ten days; instead, the u band observations from VOICE do not have a specific cadence. 
For our variability analysis, we chose to use the $r$ band data, due to the better temporal sampling.

   \begin{table}\caption{Observations in the r band used in the present work. Columns: (1): observing block i.d. (OB) of each epoch; (2): date of observation of each epoch; (3): target field; (4): total exposure time; (5): median seeing of the epoch.}\label{tdata}
\begin{tabular}{c c c c c}        

%Observation 
 OB & date & Field & $\rm T_{exp}$ & FWHM \\
(1) & (2) & (3) & (4) & (5) \\
 & y-m-d & & s & arcsec \\
\hline\hline

 703642    &  2012-08-10   & CDFS1 &  1800.0  &   0.806\\
 703646    &  2012-08-13   & CDFS1 &  1800.0  &   0.693\\
 703664    &  2012-09-02   & CDFS1 &  1800.0  &   1.019\\
 703674    &  2012-09-08   & CDFS1 &  1800.0  &   0.998\\
 703682    &  2012-09-14   & CDFS1 &  1800.0  &   0.550\\
 703686    &  2012-09-17   & CDFS1 &  1800.0  &   1.062\\
 703690    &  2012-09-20   & CDFS1 &  1800.0  &   0.874\\
 703694    &  2012-09-22   & CDFS1 &  1800.0  &   0.892\\
 779894    &  2012-10-07   & CDFS1 &  1800.0  &   0.927\\
 779919    &  2012-10-08   & CDFS1 &  1800.0  &   0.930\\
 779898    &  2012-10-11   & CDFS1 &  1800.0  &   0.921\\
 779902    &  2012-10-15   & CDFS1 &  1800.0  &   1.034\\
 779906    &  2012-10-17   & CDFS1 &  1800.0  &   0.924\\
 779910    &  2012-10-21   & CDFS1 &  1800.0  &   0.508\\
 779914    &  2012-10-25   & CDFS1 &  1800.0  &   0.862\\
 843920    &  2012-11-04   & CDFS1 &  1800.0  &   0.671\\
 843925    &  2012-11-06   & CDFS1 &  1800.0  &   0.832\\
 843930    &  2012-11-08   & CDFS1 &  1800.0  &   0.884\\
 843935    &  2012-11-10   & CDFS1 &  1800.0  &   0.762\\
 843945    &  2012-11-19   & CDFS1 &  1800.0  &   0.658\\
 779831    &  2012-12-03   & CDFS1 &  1800.0  &   0.708\\
 779836    &  2012-12-07   & CDFS1 &  1800.0  &   0.812\\
 779846    &  2012-12-13   & CDFS1 &  1800.0  &   0.545\\
 779856    &  2012-12-20   & CDFS1 &  1800.0  &   0.963\\
 779862    &  2013-01-03   & CDFS1 &  1800.0  &   0.684\\
 779867    &  2013-01-06   & CDFS1 &  1800.0  &   0.913\\
 779872    &  2013-01-10   & CDFS1 &  1800.0  &   0.888\\
 stacked    &    & CDFS1 &  18360.0  &   0.671\\
\hline\hline
606786    &  2011-10-20   & CDFS2 &  1800.0  &   1.173\\
 606792    &  2011-10-25   & CDFS2 &  1800.0  &   0.560\\
 606795    &  2011-10-28   & CDFS2 &  1800.0  &   0.918\\
 606798    &  2011-10-30   & CDFS2 &  1800.0  &   1.064\\
 606801    &  2011-11-02   & CDFS2 &  1800.0  &   0.779\\
 606804    &  2011-11-04   & CDFS2 &  1800.0  &   0.616\\
 606808    &  2011-11-15   & CDFS2 &  1800.0  &   0.607\\
 606811    &  2011-11-18   & CDFS2 &  1800.0  &   0.897\\
 606814    &  2011-11-21   & CDFS2 &  1800.0  &   0.680\\
 606817    &  2011-11-23   & CDFS2 &  1800.0  &   0.903\\
 606820    &  2011-11-26   & CDFS2 &  1800.0  &   0.638\\
 606823    &  2011-11-28   & CDFS2 &  1800.0  &   1.043\\
 606826    &  2011-12-01   & CDFS2 &  1800.0  &   0.824\\
 606829    &  2011-12-03   & CDFS2 &  1800.0  &   0.523\\
 606723    &  2011-12-14   & CDFS2 &  2160.0  &   0.883\\
 606727    &  2011-12-17   & CDFS2 &  1800.0  &   0.880\\
 606756    &  2012-01-14   & CDFS2 &  1800.0  &   0.769\\
 606760    &  2012-01-18   & CDFS2 &  1800.0  &   0.574\\
 606764    &  2012-01-20   & CDFS2 &  1800.0  &   1.003\\
 606768    &  2012-01-23   & CDFS2 &  1800.0  &   0.594\\
 606772    &  2012-01-25   & CDFS2 &  1800.0  &   0.901\\
 606776    &  2012-01-29   & CDFS2 &  1800.0  &   0.666\\
 stacked    &     & CDFS2 &  19440.0  &   0.637\\

\end{tabular}
\footnote{Columns: (1):observation block; (2): date of observation; (3): target field; (4): exposure time; (5): average seeing of the observation}
\end{table}

As described in detail in \cite{decicco2014}, the data reduction was performed by using VST-Tube, a pipeline specifically designed to reduce VST/Omegacam data \citep{grado2012}, which takes care of %  dither combination, astrometric registration, exposure and vignetting correction as well as of photometric calibration.
 the
instrumental signature removal, including overscan, bias and flat-
field correction, CCD gain harmonisation and illumination correction (IC), co-addition of the exposures for each epoch, and finally astrometric and photometric calibration of the data.

We note that in the case of wide field images, the final photometric accuracy across the entire FOV depends on the quality of the gain harmonization across the multiple CCDs, as well as on the illumination and scattered light correction, which are position and time dependent. 
The absolute photometric calibration was computed on the
photometric night 2011 Dec 17 by comparing the observed magnitude of stars in photometric standard fields with SDSS DR8 magnitudes \citep{eisenstein2011}. 
The extinction coefficient was taken from the extinction curve M.OMEGACAM.2011-12-01T16:15:04.474 provided by ESO. 
The relative photometric correction among the
exposures was obtained by minimising the quadratic sum of
differences in magnitude of the multiple detections.
The dependence of such corrections on the position within the FOV is taken care of during the gain harmonisation and IC correction phase mentioned earlier.
We had a total of 29 epochs for the CDFS1 and 22 for the CDFS2 (see Table \ref{tdata} and Fig. 1).
We retained only those observations with good seeing, i.e. FWHM $<$1.2 arcsec, in order to optimise the signal-to-noise ratio (S/N) of the AGN against the host galaxy light, while minimising source blending and position uncertainties.
For this reason we excluded the CDFS1 observations made on 2012 Sept 5 (OB=703670) and on 2012 Sept 24 (OB=703698) with seeings of 1.28 and 1.44 arcsec, respectively, and the CDFS2 observation made on 2012 Feb2 with seeing 1.46 arcsec (OB=606780).

As we see in Sect. \ref{discussion}, we also exploited a number
of overlapping surveys to validate our catalogue of
variable objects: SWIRE \citep{lonsdale2004}
%providing Spitzer imaging in 7 IRAC and MIPS IR bands,
 and
SERVS \citep{mauduit2012}. %providing deeper Spitzer imaging in IRAC 3.6 and 4.5 $\mu$ channels.  
For both SWIRE and SERVS, we used the Spitzer Data Fusion catalogues
\citep{vaccari2010} \footnote{http://mattiavaccari.net/df/}.
  These
provide a deeper source extraction than the publicly available SWIRE
DR5 \footnote{http://irsa.ipac.caltech.edu/data/SPITZER/SWIRE/} and SERVS DR1 \footnote{http://irsa.ipac.caltech.edu/data/SPITZER/SERVS/} catalogues, as well as a wealth of multi-wavelength ancillary
data. Over the area explored in this paper, data are available in  3.6, 4.5, 5.6, 8, 24, 70, and 160 $\mu m$ bands, and in the U, g, r, i, and z filters.
We also used the spectral energy distribution (SED) information from the SWIRE
photometric redshift catalogue presented in
\cite{rowan-robinson2013}.

Finally, we used the photometric redshift catalogue and  SED (in the 0.2 to 10 $\mu$m observed-frame range) classification published in \cite{hsu2014} for sources in the ECDFS area. 
The area explored in \cite{hsu2014} is the one covered by the Multi-wavelength Survey by Yale-Chile (MUSYC) coverage \citep{gawiser2006,cardamone2010}, which encloses about one-fourth of the CDFS1 area studied in the present work.
Inside that area, \cite{hsu2014} have used  the information provided by other campaigns: photometric data from TENIS by \cite{hsieh2012} and CANDELS by \cite{guo2013}; X-ray data by \cite{xue2011,rangel2013,lehmer2005,virani2006}.
More details on the data used in \cite{hsu2014} are provided in Sect. 5.2.
Figure 1 shows the overlap between the VST CDFS
studied in this work and the SWIRE, SERVS, and ECDFS surveys just described.

\begin{figure}
\includegraphics[angle=0,width=9.cm]{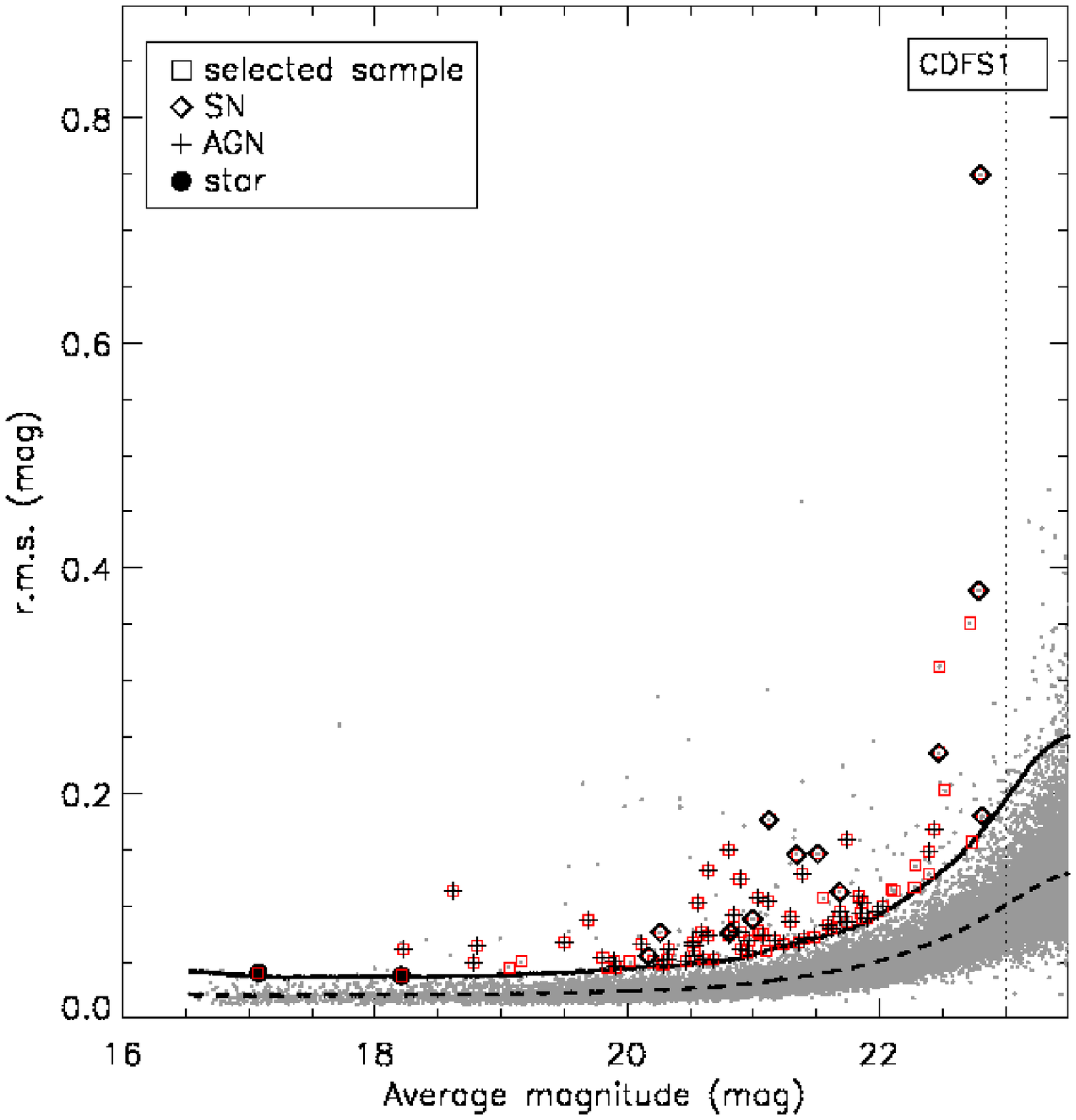}\\
\includegraphics[angle=0,width=9.cm]{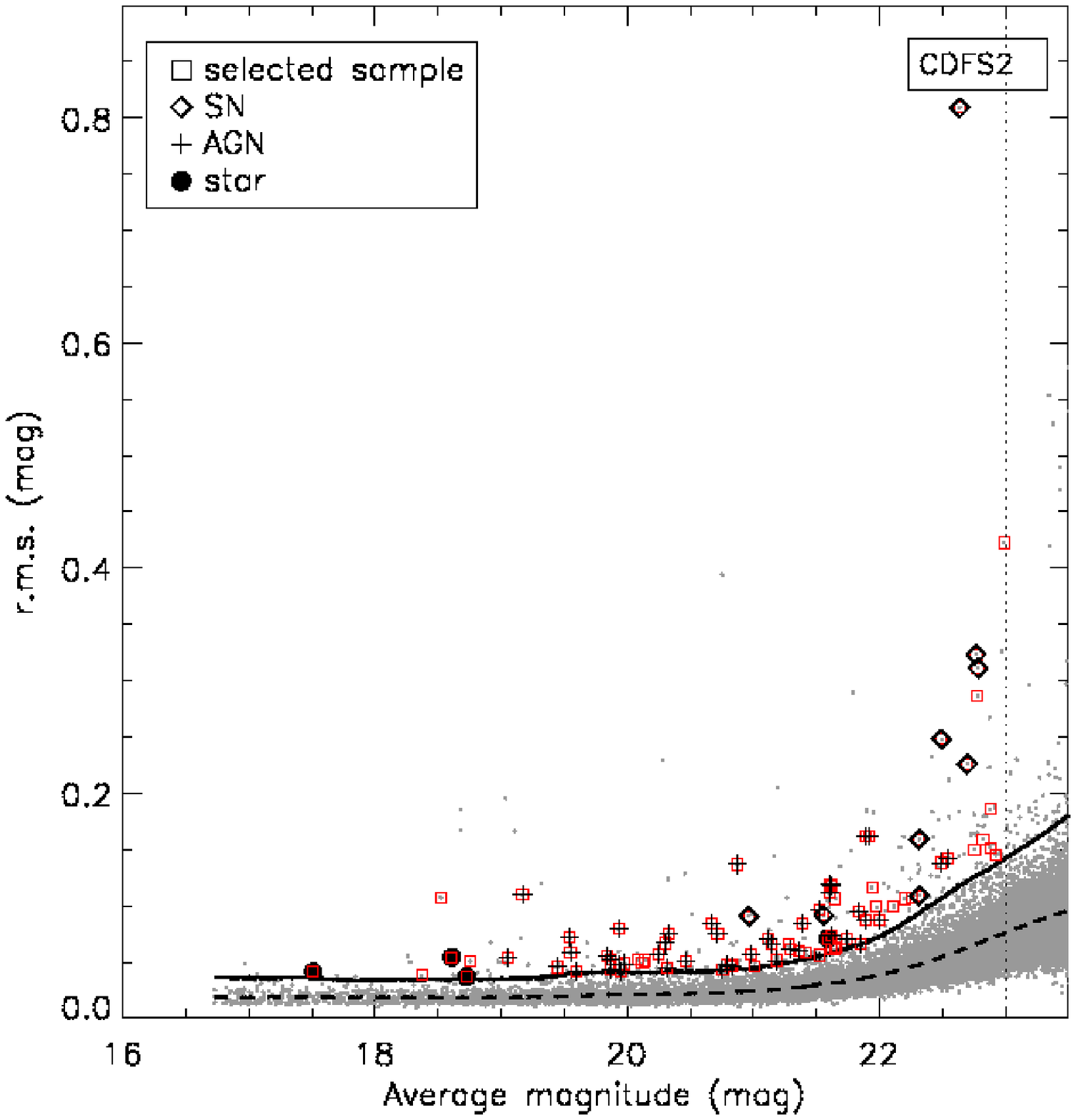}
      \caption{R.m.s versus average magnitude of the full masked samples of CDFS1 (top panel) and CDFS2 (bottom panel). Dashed line: running average of the r.m.s.; solid line: 3$\sigma$ variability threshold adopted to select variable sources (see text). Sources reported in Table 3 are indicated with the following symbols: squares: all sources of the selected sample; crosses: AGN; circles: stars; diamonds: SNe (as discussed in Sect. 5)
              }
         \label{fig:selection}
   \end{figure}

\section{Catalogue extraction}\label{method}

\begin{figure}
\includegraphics[angle=0,width=7cm]{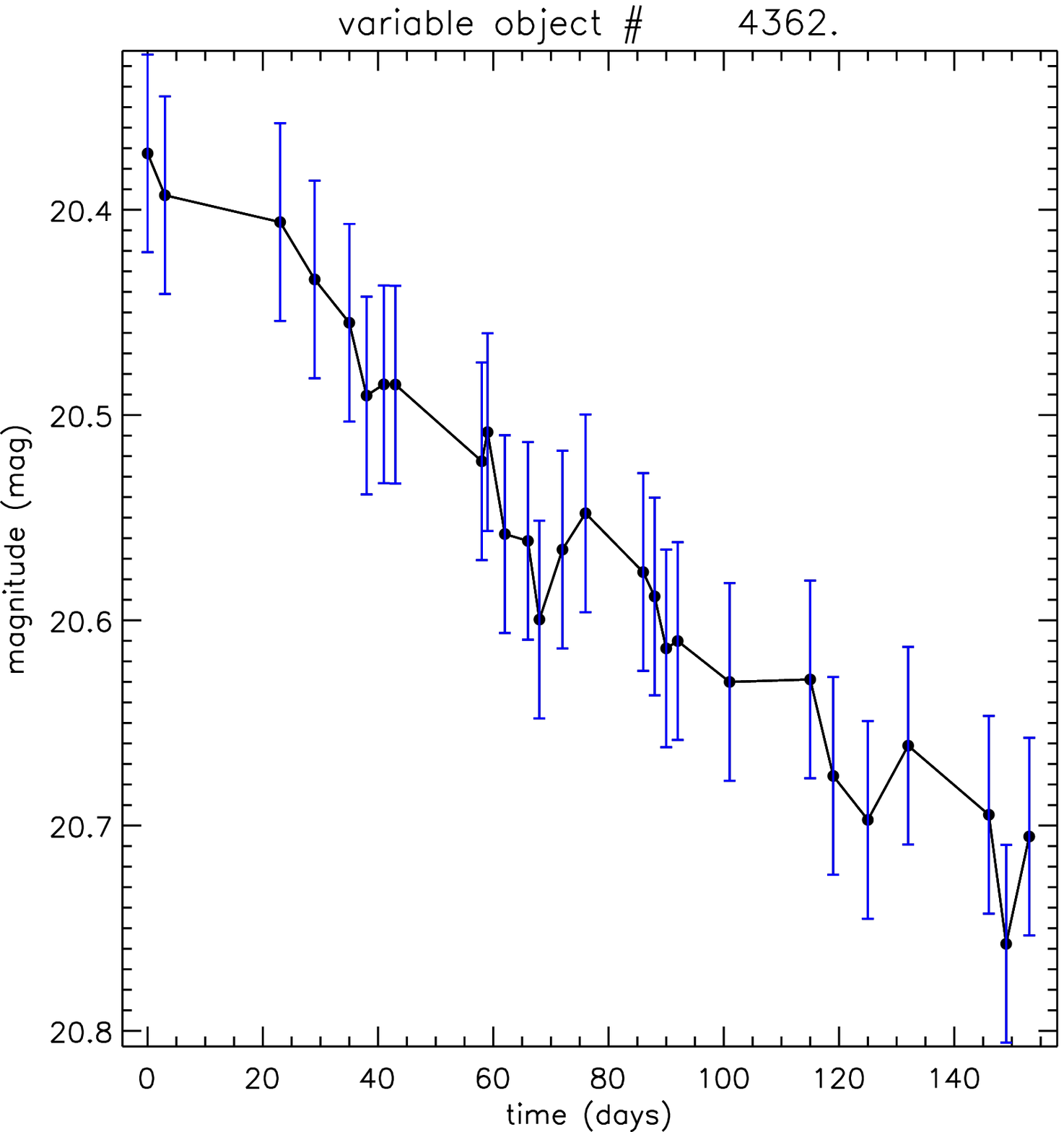}\\
\includegraphics[angle=0,width=7cm]{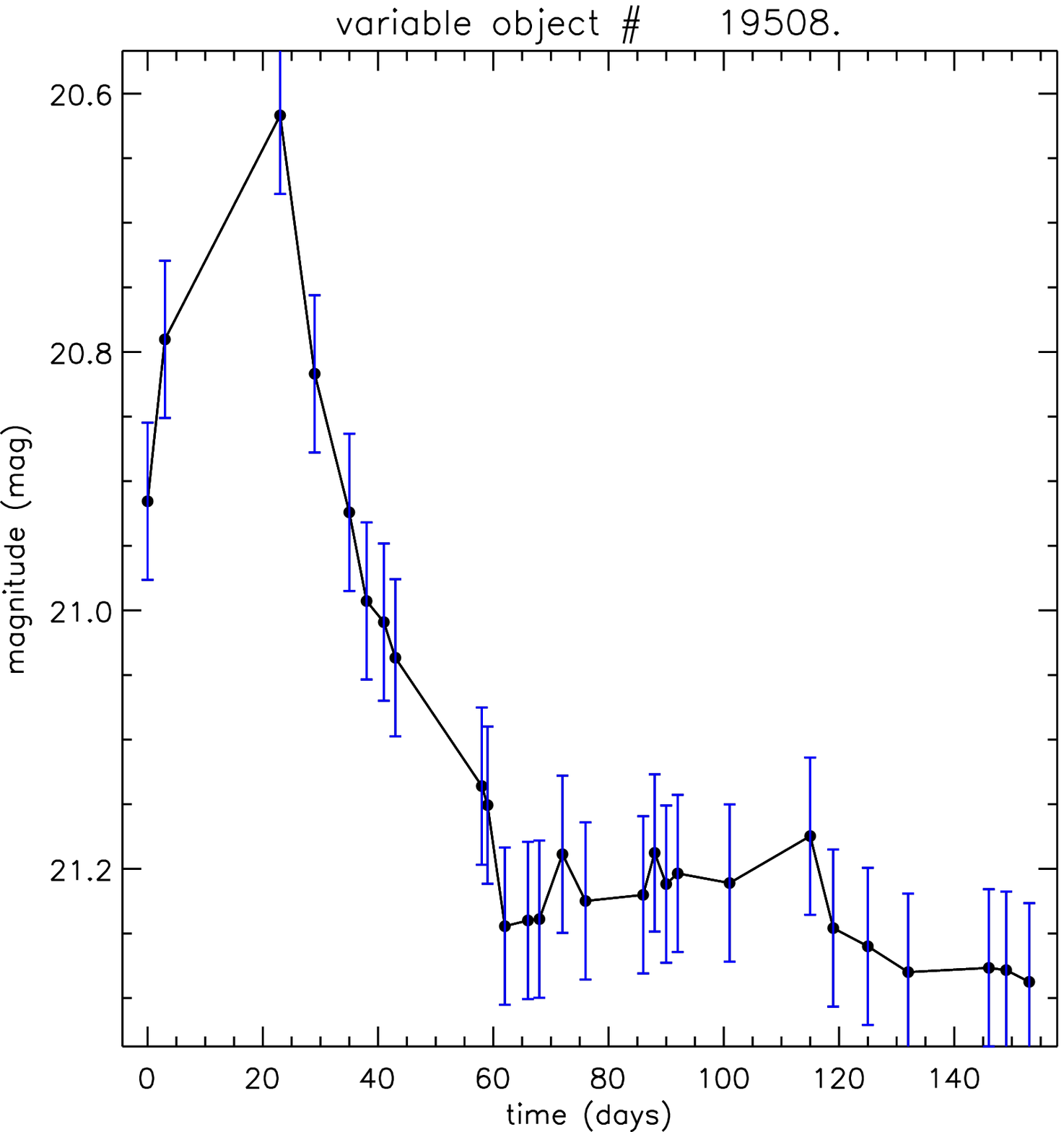}
      \caption{Top panel: lightcurve of an AGN candidate (ID 4362). Bottom panel: lightcurve of a supernova (ID 19508). The error bars do not represent the formal photometric error except the $\langle \sigma \rangle  +3 \times rms_{\langle \sigma \rangle }$ variability limit and are shown to allow the reader to visualise the significance of the variability. } 
         \label{agn}
   \end{figure}

For detecting variable sources, we used the method presented in \cite{trevese2008}, which was adapted to work on VST data as described in Paper I. While referring to these papers for details, we summarise the main steps of the procedure here: extraction of the catalogues, aperture correction, masking, and construction of the master catalogue.

For the first step, we used Sextractor to produce source catalogues for each epoch and to measure the photometry in a set of concentric apertures in order to measure the isophotal, aperture, and total source magnitudes. 
The aperture size for AGN identification should
include the majority of the flux from the central source and minimise the
contamination from nearby objects or the host galaxy. 
To build the lightcurve of each source we used the flux measured within a 2 arcsec diameter aperture, centred on the source
centroid, which corresponds to about 70\% of the flux from
a point-like source given our average seeing. The 2 arcsec diameter is twice the point spread function (PSF) full width at half maximum (FWHM) for the worst images in our dataset (as can be seen in Table 1),
allowing to optimise the S/N on the aperture flux as shown by, for example, \cite{pritchet1981}.
%% Based on our tests on the VST images, the 2 arcsec aperture also represents 
%% twice the source half-flux radius, making sure that we are retrieving a significant fraction of the source total flux and thus minimising the uncertainties on the total extrapolated flux. 

To take the effect of variable seeing into account, we applied aperture corrections to each individual epoch, making use of the growth curves derived from bright stars. The stars were chosen to have circular isophotes, isolated from other sources, and have a low local background.
We verified that the PSF is sufficiently uniform across the field of view and that the choice of one specific star does not introduce significant differences.
We then corrected all our measurements to 90\% of the total flux\footnote{This choice is arbitrary and is retained for consistency with Paper 1. In fact, the exact value to which we correct is irrelevant for the purpose of variability detection since it is only intended to compensate for seeing effects, thereby normalising all measurements to the same fraction of the total flux.}.

The correction technique applied here is based on the assumption that the source is point-like, so
it has the effect to over correct extended sources containing faint AGNs %(most of which have been excluded in the present work, since we focused on sources with $r<$23 mag, see Sect. 4)
 because it improperly
corrects part of the flux from the galaxy. 
However, as discussed in Paper I, this effect increases the average variability of the most extended sources by less than 0.01 mag in total r.m.s..
  The result is that our fixed variability threshold returns a sample somewhat biased towards more extended sources. However, note that the 0.01 mag variability increase is an upper limit, applying only to the most extended sources; in fact, the majority of the variable candidates found in this work are point-like.

The limit magnitude of the shallowest epoch is r(AB) $\sim$ 24.5 mag for a source detected at the 5$\sigma$ significance level. The completeness magnitude (of the shallowest epoch) is r(AB) $\sim$ 23 mag at the 90\% confidence level, therefore we focus on the variable sources that are more luminous than 23 mag (see Sect. 4).
For each epoch, we masked out regions including halos of bright stars, spikes, artefacts, reflection, and residual satellite tracks. A code developed by \cite{huang2011A} was used to produce, for each epoch, specific  pixel masks for halos and spikes due to bright stars, taking the star position on the frame and telescope orientation into account. We manually added more pixel masking for extended regions corresponding to other artefacts (i.e. satellites tracks and borders). After the masking procedure, the total number of sources in the master catalogues decreases by about 20 \%: we note that in this step we prefer to use a conservative approach to producing a clean sample that favours its purity at the expense of its completeness.
We matched the source catalogues of each epoch using a matching radius of 1 arcsec to obtain a single global catalogue to be used for our variability analysis.
%% These catalogues include 202723 sources in the CDFS1 and 128175 in the CDFS2, including transients.

Since we aim at building a robust catalogue of variable but persistent sources, we selected only the sources with detections in at least six epochs (out of 27 and 22 epochs in the CDFS1 and CDFS2, respectively). %,thus excluding non-recurring fast transients. 
The selection  (r$<$23 mag and $N_{epochs}\geq 6$) yields 18~954 in the CDFS1 and 14~340 in the CDFS2.
The different number of sources in the two fields is due to the different masked areas and, to a minor extent, to an average lower completeness magnitude (amongst all epochs) in the CDFS2.
% This selection yields 45776 and 35120 sources in the CDFS1 and CDFS2, respectively. These are 80896 in total (in the CDFS1 plus CDFS2).
 
\section{Selection of variable sources}\label{results}
As in Paper I, we determined the average magnitude and the r.m.s. of the lightcurves (LC) for each source. 
We then computed the running average of the individual r.m.s. $\langle \sigma_i^{LC} \rangle $ and its own standard deviation $rms_{(\sigma_i^{LC})}$ over a 0.5 mag-wide bin.
We defined a source as variable if it satisfies the condition:
$\sigma_i^{LC} > \langle \sigma_i^{LC} \rangle  +3 \times rms_{(\sigma_i^{LC})}$.
%With this criterium, we found 532 variables in the CDFS1 and 481 in the CDFS2. 
Figure \ref{fig:selection} shows the running average and the standard deviation as a function of the average source magnitude.  The sources above the solid line are flagged as variable candidates.
 Our approach closely follows the one adopted in other works \citep{trevese2008,bershady1998}.  Indeed this method does not allow an a priori statistical estimate of  the rate of false positives (as, for
instance, in \cite{villforth2010} using HST data), but this is because we do not know the statistical distribution of our errors (at odds with
the much better characterised uncertainties affecting space data, such as those used in \cite{villforth2010}, for instance). While these are
dominated by Poissonian uncertainties in the faint regime, at brighter magnitudes the photometric uncertainty is limited by the calibration accuracy of the data (Sect. 2). Thus our approach relies on the definition of a lightcurve r.m.s. threshold, and we investigate the nature of the candidates by comparing to optical, IR, X-ray, and spectroscopic surveys (Sect. \ref{discussion}). 
%% In any case the sample of variable candidates represent $\sim0.7$ \% of the population, i.e. more than expected above a 3 $\sigma$ detection limit for a normal distribution.

To remove sources whose variability may be spurious, we visually inspected each candidate.
We retained only those candidates not affected by evident artefacts (e.g. residual stellar diffraction spikes, satellite tracks, strong background gradients, etc.) and/or close neighbours (with distance within 2 arcsec from the source and magnitude lower than mag(source)+1.5 as in \cite{decicco2014}. For these sources the variability could be due to the contamination of the nearby source) and/or centring issues (for example in very extended sources with irregular profiles for which the identification of the centroid could be problematic). 
The sample of variable sources
 is made of 113
candidates in the CDFS1 and 122 candidates in the CDFS2, i.e. 235 candidates in total. %% : these numbers represent ~0.7\% of the sources in the catalogue (detected in more than 6 epochs and with $r$ $<$ 23 mag), i.e. more than expected above a $3\sigma$ detection limit for a normal distribution
To each candidate we attributed a quality flag ranging from 1 to 3, as in Paper I: 

\begin{itemize}
\item Flag 1: the candidate is robust, with  no evident photometric or aesthetic problems (60~\% of the 235 candidates);
\item Flag 2: the candidate is likely to be variable, and the photometry may be slightly affected by the presence of a nearby companion or by minor aesthetic defects (e.g. faint satellite tracks) (15~\%);
\item Flag 3: the candidate is very likely spurious, and its variability is uncertain (the remaining 25~\%).
\end{itemize}
We retained only the candidates with Flags 1 and 2 and
obtained a sample of 175 sources (hereafter `selected sample'), including CDFS1 and CDFS2 sources listed in Table \ref{tresults}.
We point out that the original criterion for choosing objects with more than six epochs was devised to explore all types of sources included in our survey, including the transients. In principle, lightcurves with different numbers of epochs and S/N may have different r.m.s. distributions, thus yielding a biased sample. 
However, we note that 97~\% of the sources in our catalogues lack at most three epochs. Moreover, only two sources selected as variable lack more than five epochs (in Table 3, they are the sources with ID: 29449, 290) and both are confirmed SNe from the SUDARE-I SN search (Sect. 5.3).

 Figure \ref{agn} shows the lightcurves of two sources to illustrate the phenomenology encompassed by this study.
 The top panel of Fig. \ref{agn} displays a candidate whose AGN nature is supported by several diagnostics that will be discussed in Sect. \ref{discussion} and reported in Table \ref{tresults}, while the bottom panel of Fig. \ref{agn} shows a supernova candidate.

%
%                                                One column figure
%----------------------------------------------------------- S_vib
  \section{Characterisation of variable sources}\label{discussion}

In this section, we validate our variable sources by comparison with SN catalogues, X-ray and IR data, and  SED information. The main purpose is to assess the purity and the completeness of our sample of 175 candidates and to compare it with the results obtained in Paper I, using additional diagnostics with respect to our previous work. 

Table \ref{tresults} summarises the properties of the
selected sample of 175 variable sources. The columns refer to properties, including AGN/non-AGN indicators, discussed in the following sections.

\subsection{Contamination by supernovae} 

In this section, we consider and quantify the contamination by SNe in our final sample of 175 variable sources. On the basis of the results derived in Paper I for the COSMOS field, we expect that
the variability-selected sample contains a fair fraction of SNe ($\sim$14$\pm$4 \%). The visual (qualitative) inspection of the lightcurves allowed us to identify 24 (out of
175) sources as likely SNe. (They display a rapid luminosity increase,
followed by a steep decline and a plateau, bottom panel of Fig. 3.)

To validate this identification, we compared our classification with the results of the SN search team (Cappellaro et al., in prep., hereafter SUDARE-I), which adopts a quantitative approach to detecting SNe using the SUDARE-VOICE data. They apply an image subtraction method with respect to a reference epoch, after degrading the images to the lowest resolution image to select variable sources. SUDARE-I selects the sources identified as SNe based on lightcurve fitting in three bands (g, r, i) using a grid of SN templates, where the SN type, the phase, and the reddening are free parameters, and assuming the photometric redshift of the host galaxy as a prior. In this respect the SUDARE-I approach is more general than our visual inspection since it does not rely on the detection of a rising phase, because the free phase parameter and LC fitting allows SNe detected after the peak to be identified. The SN templates used in the fitting are Ia; II, IIn, Ib, and Ic.  An estimate of the reddening for the host galaxies is obtained from SED fitting and is applied to the full lightcurve, but we fitted the reddening of the SN independently of the reddening of the host. 

The photometric redshift of the host galaxy is derived through SED fitting, using the Eazy code \citep{brammer2008}, which fits the galaxy SED with a linear combination of synthetic templates. The default template set is optimised to match semi-analytic galaxy formation and evolution models that are complete to very faint magnitudes, rather than magnitude-limited spectroscopic samples
 (see \cite{brammer2008} for details). The SED fitting is based on a set of 8-12 broad-band filters\footnote{The multi-wavelength coverage depends mainly on the source flux since u and IR data described in Sect. 2 are not as deep as the optical ones.} Spectroscopic redshifts are only available for a subsample of galaxies and the collaboration obtained spectra through dedicated observing time for 17 SN candidates. Although the analysis of the SN data is still in progress (and will be presented in Cappellaro et al. in preparation), we find that the measured SN rate is consistent with published estimates in the literature, within
the errors.

We cross-correlated our variable candidates in the selected sample with the list of sources discovered in SUDARE-I. 
Twenty-one sources, all belonging to our SN candidates, are classified as SN in SUDARE-I as well. % and the remaining 78 have been generically flagge as non-SN.
  %(amongst the remaining common sources there are 65 AGN and 13 sources classified generically as non-SN and non-AGN). 
There are three additional sources (the ID are 44289, 27638, 105093) with lightcurves similar to those found in SN  (as explained before, their lightcurves display a steep increase,
followed by a steep decline and a plateau, such as the one shown in
Figure 3, lower panel), but not confirmed as SN in SUDARE-I because the adopted SN templates did not fit their lightcurves well. As we see later, we could not find other multi-band data to use as diagnostics for two
of them (105093 and 44289). For one source (27638), one diagnostic
instead supports its AGN nature.
No additional SN has been found by SUDARE-I within our sample of r$<$23 sources, indicating that our visual inspection was able to identify all bright SN candidates.

In summary, the fraction of SN listed in the selected sample is 12$\pm2$ \% (which corresponds to the 21/175 SN found in the selected sample. This fraction agrees, within the error bars (95 \% confidence level), with the results obtained from the COSMOS field analysed in Paper I.

\subsection{SED and X-ray detections}

For characterising the variable sources and the validation
of the AGN catalogue, it is necessary to use other diagnostics. Unfortunately, broad SED coverage, optical spectroscopy, or
X-ray observations is only available for a small fraction of the survey
area: while the ECDFS has been covered by a variety of multi-band and
spectroscopic campaigns, the remaining area currently has patchy
coverage.
We used the information contained in \cite{hsu2014} to extract X-ray and SED data for our variable candidates located within the ECDFS area, which covers one-eighth of the full region explored in the present paper. 
The authors  computed photometric redshifts via SED fitting for all the galaxies detected in the GOODS/CANDELS area using the deep ($\sim$ 29 mag), space-based PSF fitted photometry, presented in \cite{guo2013}. \cite{hsu2014} also use UV photometry from GALEX and the PSF fitted intermediate-band photometry from Subaru, useful for determining the presence of
emission lines. For the area outside the CANDELS region, the photometry presented in MUSYC \citep{gawiser2006,cardamone2010}, covering optical, NIR, and MIR was used, in
addition to deeper IR data from the TENIS survey \citep{hsieh2012} and GALEX data.
For the X-ray band, \cite{hsu2014} collect information from the 4 Ms source catalogues by \cite{xue2011} and \cite{rangel2013}, covering the central part of the CDFS area and reaching sensitivity limits of 3.2 $\times$ $10^{-17}$, 9.1 $\times$  $10^{-18}$, and 5.5 $\times$  $10^{-17}$ erg/$\rm{cm}^2 $/s in the full (0.5-8 keV), soft (0.5-2 keV), and hard (2-8 keV) bands, respectively. The data for the enlarged ECDFS area are instead collected from the 250 ks catalogues of \cite{lehmer2005} and \cite{virani2006},  with sensitivity limits of 1.1 $\times$ $10^{-16}$ in the soft band and 3.9 $\times$ $10^{-16}$ erg/$\rm{cm}^2 $/s in the hard band.
Photometric redshifts were computed using tuned procedures able to disentangle AGN from stars and galaxies where the X-ray luminosity is produced by stars (see \cite{hsu2014} for details).

  There are only 15 sources in common between the sample presented in \cite{hsu2014} and our selected sample, using a matching radius of 1 arcsec.
 The 15
common sources belong to the CDFS1, which encloses the ECDFS. 
  Twelve of the 15 common sources are detected in the X-rays with luminosities and optical/X-ray flux ratios typical of both Types I and II AGNs ($-1< log (f_{opt}/f_X)<1$ (see e.g. \citet{mainieri2002} and \citealt{xue2011}) and in agreement with Paper I. Moreover, their SEDs require a strong AGN contribution (in particular in the NIR part of the spectrum, as can be seen in the example shown in the top panel of Fig. 4).
 All these sources have also been identified as non-SN on the basis of the inspection of their lightcurves in Sect. 5.1. 
The remaining three sources are undetected in the X-rays (ID: 11324, 5854 and 15811) and their best-fit SED template shows no evidence of a significant AGN contribution. (An SED example is shown in the bottom panel of Fig. 4.) These three sources were identified as SN according to their lightcurves in Sect. 5.1. Therefore, we conclude that they are SN explosions in normal galaxies.
In summary, our variable candidates  with a counterpart in the ECDFS region, are either confirmed to contain an AGN according to their broadband properties, or alternatively, when no AGN signature is detected, their lightcurves are consistent with SN.

\begin{figure}
\includegraphics[angle=0,width=9cm]{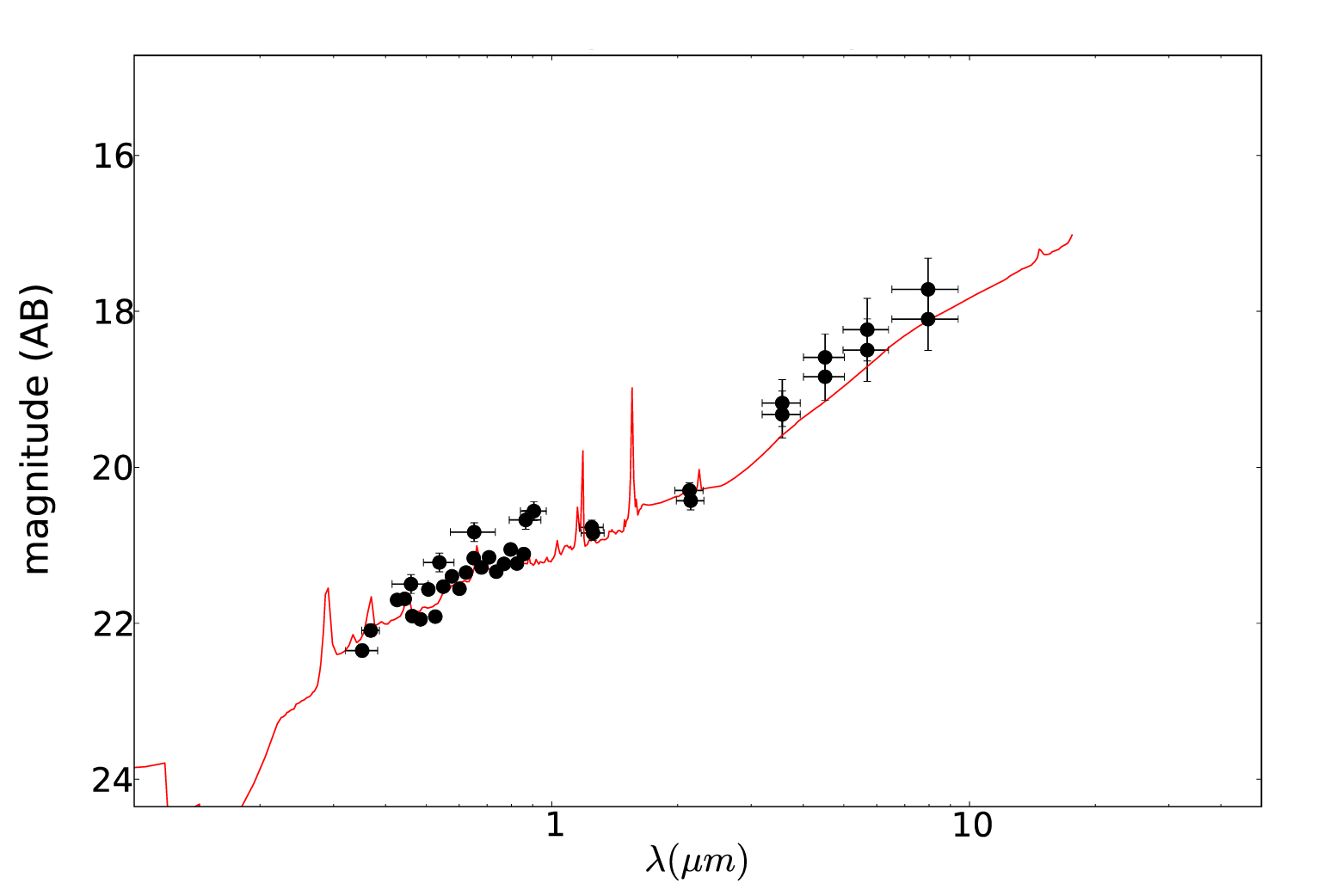}\\
\includegraphics[angle=0,width=9cm]{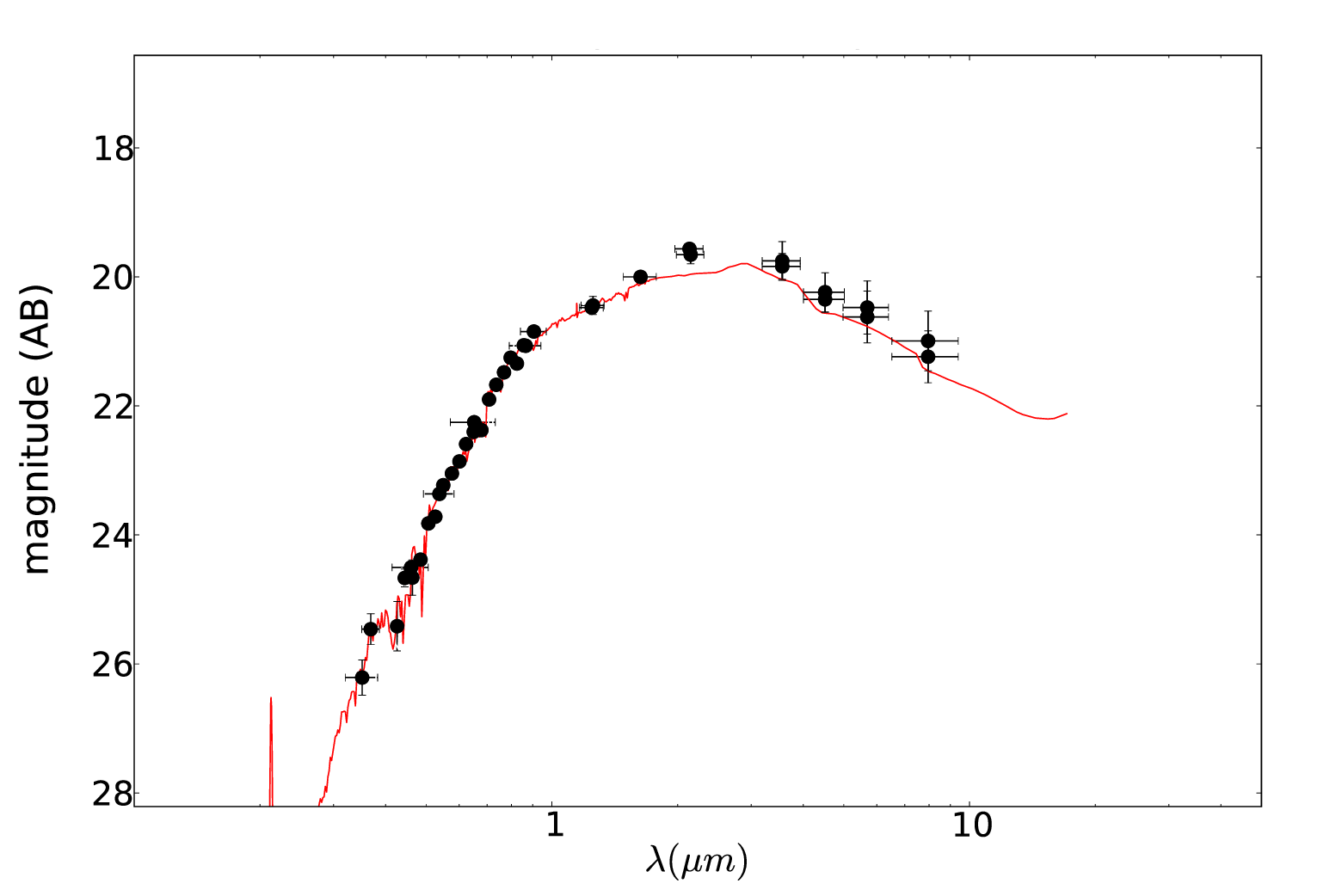}
      \caption{SED of two sources by \cite{hsu2014}. Top panel: source 2663 (an AGN candidate, see Table 3); Bottom panel: source 11324 (a SN candidate).}
         \label{sed_agn}
   \end{figure}

%\subsubsection{Optical spectroscopy}
 Two examples of \cite{hsu2014}  fitting are presented for source IDs 2663 and 14321 in our sample (the first one X-ray detected and the second one not). The plots show the
photometry, expressed in AB magnitude and in observed frame, with the best-fitting template. Additionally, nine AGN candidates have optical spectroscopy available from \cite{boutsia2009} and were identified as broad-line AGN. %% Col. 10 of Table 3 flags as 1 these nine AGN with optical spectroscopy confirming their AGN nature. 
As we can see from Table 3, the X-ray detections and SED fitting just discussed \citep{hsu2014} confirm this identification.

\subsection{IR catalogues (SERVS+SWIRE)} 

For further discussion and validation of our catalogue, we used the SERVS \citep{mauduit2012} and SWIRE \citep{lonsdale2004} samples to exploit the four IRAC bands (the fluxes, respectively, at: 3.6, 4.5, 5.8, 8.0
$\mu$m) and the CTIO MOSAIC2 Ugriz optical imaging available through the Spitzer Data Fusion over our VST survey.
SERVS and SWIRE overlap an area of $\sim$ 6 deg$^2$
(with 281149 common sources, constituting the catalogue called hereafter SERVS+SWIRE).
Almost all the sources in the selected sample fall in the area of SERVS+SWIRE  (172 out of 175), and we could find the corresponding IR sources for 158 of them, within a radius of 1 arcsec.
%(three of the 17 sources are very close to the border and thus their counterparts may be missing.: ID 105093, 108618, 110606) The rest is in the common area
A 1 arcsec radius is larger than the average offset ($\sim0.3$ arcsec) between the optical and IR catalogues  but was chosen to ensure that even nearby extended sources with ill-defined centroids (e.g. late-type galaxies) are properly matched. Given the average source density in our fields, we estimate $\sim 1$\% false matches.  
%This has been proved adopting different matching radii from 1 to 3 arcsecs and checking that the number of matching sources changes only of few units.
%% We tried to use a wider cross-correlation radii to check whether the number of cross-correlating sources is affected by the limited accuracy and astrometric precision of the surveys: at 1.5 arcsec we have one more matching source. The number of common sources increases to 160 if we use instead a radius of 3 arcsec. 
Among the 14 sources in the common area and without SERVS-SWIRE counterparts, there are seven SN (of those discussed in Sect. 5.1): while the transient events (the SN explosions) can be detected using our method, their hosts are likely to be normal galaxies without strong IR emission so not included in the SERVS+SWIRE catalogues.
 
For the common subset of 158 sources, we used optical and IR colour-colour diagrams to constrain their nature, as described in detail in the following sections.

\subsubsection{Optical-NIR diagnostic} 
\begin{figure}
%\centering
 \includegraphics[angle=0,width=10cm]{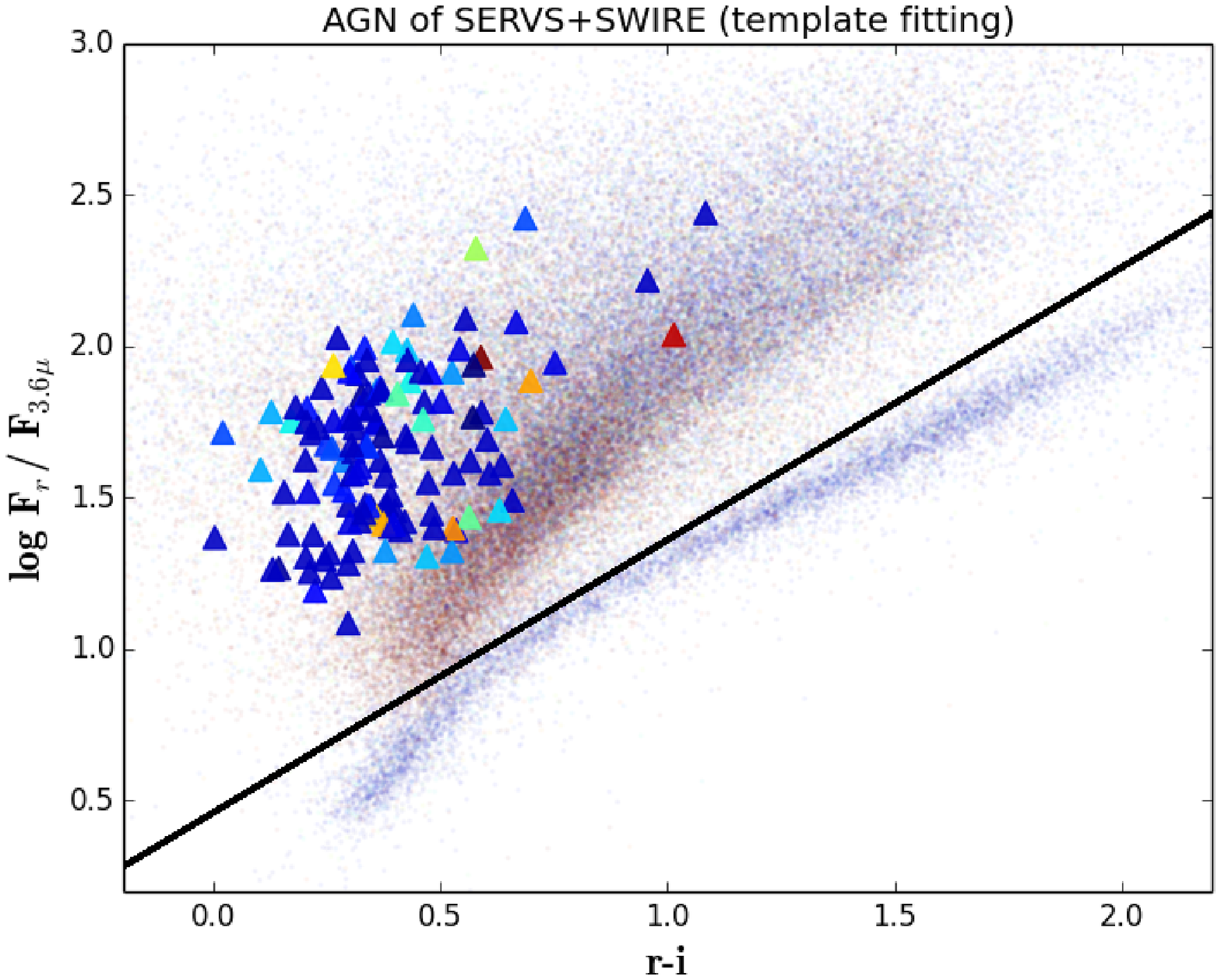}
 \includegraphics[angle=0,width=10cm]{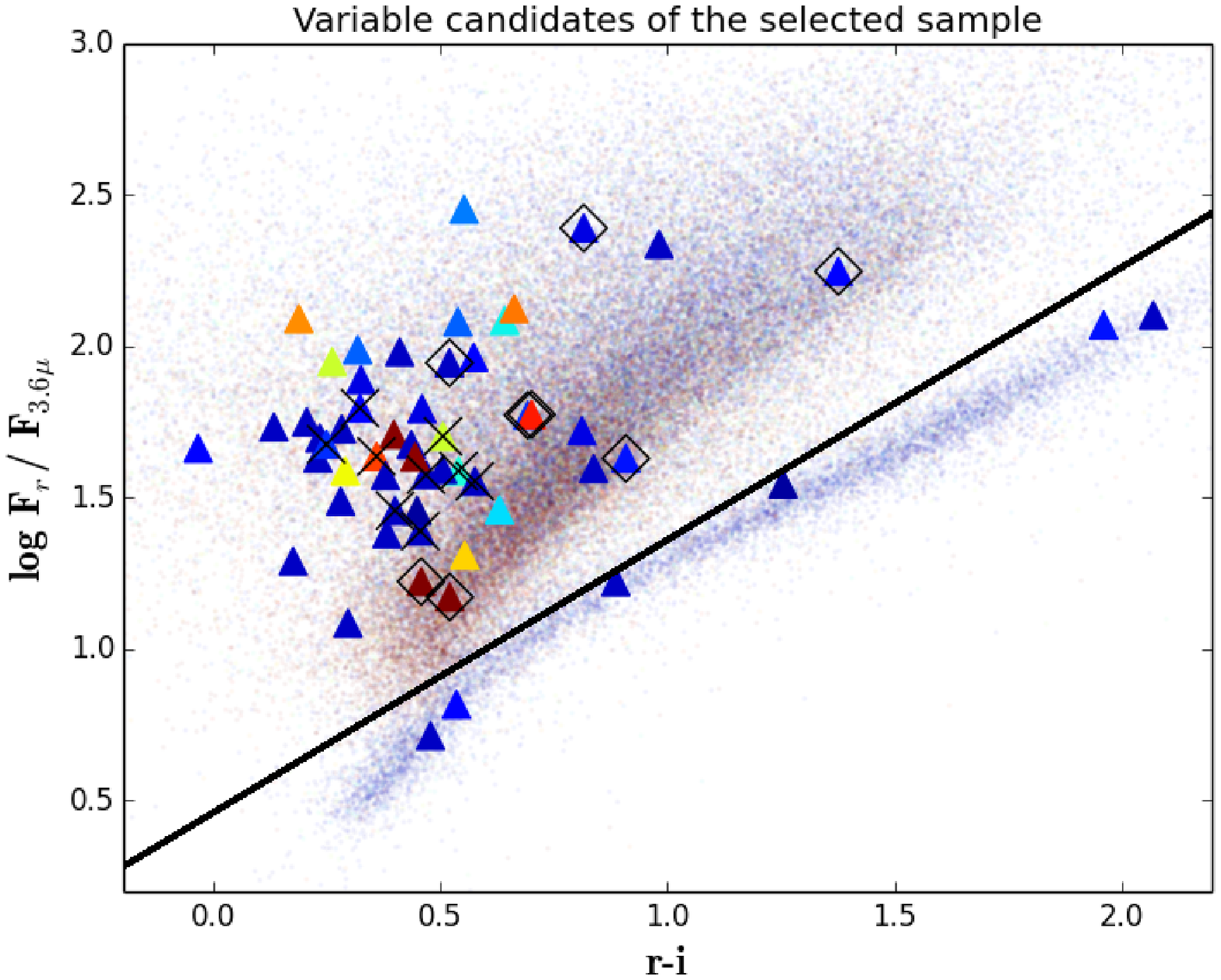}
      \caption{ Flux (F$\lambda$) ratios between the r band and 3.6 $\mu m$ versus  r-i colour.
In the \texttt{top panel}, the triangles represent sources flagged as AGN in the SERVS+SWIRE catalog, on the basis of optical/IR template fitting (133 AGN, see Sect. 5.3.1). In the \texttt{bottom panel} the triangles represent the 57 sources belonging to our ‘selected sample’ with available i, r, and 3.6$\mu$m photometry in the SERVS+SWIRE catalogues (see Sect. 5.3.1). Diamonds and crosses label the SN and X-ray detected sources, respectively (see Sect. 5.2). 
In both panels, the small points represent the whole SERVS+SWIRE control sample with available r,i, 3.6 $\mu m$ photometry (82254 sources). The colours from red (extended sources) to blue (pointlike sources) indicate the increasing stellarity (defined in footnote 9) as measured from the VST images for our ‘selected sample’ or extracted from the SERVS+SWIRE catalogues in the optical r band for the control sample. The solid line divides the plane into the stellar region and the non-stellar region.
              }
         \label{36ri}
   \end{figure}
In Fig. \ref{36ri} (bottom panel) we compare the $r-i$ colour versus the 3.6 $\mu m$ to $r$ band flux ratio of our variable candidates with the SERVS+SWIRE source catalogue. This diagram has been proposed by \cite{rowan-robinson2013} to separate stars from galaxies. The analysis is limited to the 57 objects (of the selected sample) with available data in r, i, and 3.6 $\mu m$.

The populations represented in the plot are segregated into two regions: stars and extra-galactic objects. 
In Fig. 5, a line separates the two regions: below that line, the total number of stars in SERVS+SWIRE catalogue is 15 times higher than the number of extended objects.

Figure \ref{36ri} (bottom panel) shows that 6 out of 57 variable sources are along the stellar sequence, so very likely variable stars. We deduce that the fraction of variable stars is probably around 10$\pm$4~\% (6 out of 57).  
This result can be compared with that of \cite{sesar2007} who found 6.4 \% of the variable sources identified as RR Lyrae stars, and 18.3 \% identified as stars in the stellar sequence. In our sample we can exclude that the six sources in the stellar locus are RR Lyrae because they do not satisfy the selection criterion for their r-i colours: $-0.15<r-i<0.22$ \citep{sesar2007}. Moreover, the variability timescales of RR Lyrae stars are much shorter than the scales sampled in the present survey (from several hours to one day).
 The specific nature of our six star candidates is not clear.  
For one of them (ID 26458), the optical SED classification (reported in Col. 12 of Table \ref{tresults}) confirms that there is no AGN contribution.
 Our 10 \% of stars is lower than the 18 \% of objects in the stellar sequence found in \cite{sesar2007}. Since \cite{sesar2007} is based on the SDSS Stripe 82 survey located at lower galactic latitudes, it is expected that more stars were observed in that survey.

Outside the stellar sequence shown in Fig. 5 there are eight SN, confirmed through the lightcurve fitting of SUDARE-I (48377, 19508, 5854, 15811, 7120, 11324, 6690, 33242). 
There are four extended sources, with erratic lightcurves clearly different from those of SN (from our inspection and Cappellaro et al. in prep). Three of them (those with IDs 7134, 8056, 19168)
are confirmed AGNs according to the IR diagnostic plot discussed in Sect. 5.3.2, while the nature of the last source (ID 24635) is unknown.

Amongst the 51 sources with non-stellar colours, 41 have unresolved profiles according to their stellarity index \footnote{CLASS-STAR given by Sextractor, which is
the probability for a source of being point-like -from zero, for
extended sources, to one, for point-like sources-}. Although this diagram is mainly aimed at
separating stars from galaxies, 
it is visually clear that our candidates have colours that differ, on average, from the bulk of the galaxy population.
Previous studies \citep{berta2006,tasse2008} show that both AGNs and high-z galaxies populate the area where our variable sources are located. 
As a further check, we plot in Fig. 5 (top panel) the AGNs identified in the  SERVS+SWIRE catalogue on the basis of the optical/IR template fitting performed by \cite{rowan-robinson2013}. As can be seen in the bottom panel of the same figure, most of our variable sources not identified as stars have colours that areconsistent with those of AGNs. 
 The non-stellar colors and compactness of our candidates, when coupled with their variability, support the idea that these sources host AGNs, SNe or some other type of transient source. The
lightcurve inspection further allows to remove SNe, thus leaving
the AGN classification as the most likely one. This interpretation, however, needs to be validated using additional diagnostics such as those presented in Sects. 5.2 and 5.3.2.

\subsubsection{IR diagnostic} 
In this section we make use of the mid-IR colours in order to confirm the identification of our AGN candidates.
\begin{figure}
 %\centering
 %  \includegraphics[angle=0,width=9.cm]{figures/58_45_58_36.ps}
 \includegraphics[angle=0,width=10.cm]{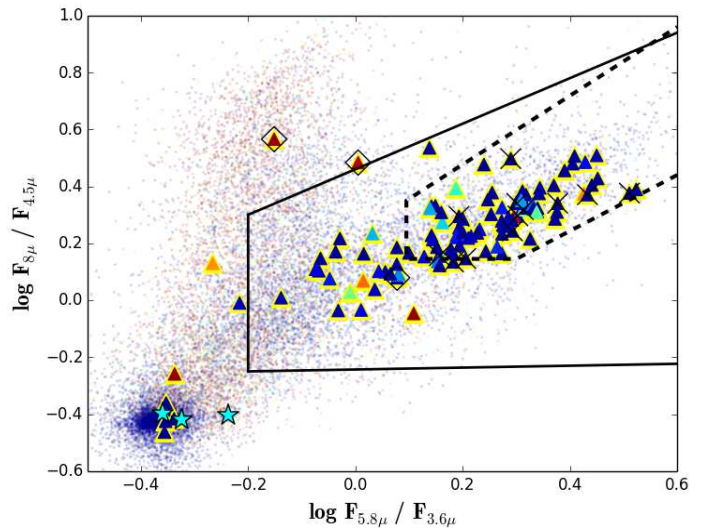}
      \caption{%Left-hand panel: Flux ratio at 5.8 and 4.5 $\mu$ versus flux ratio at 5.8 and 3.6 $\mu$ according to \cite{lacy2004} diagnostic. Right-hand panel: 
Flux (F$_{\lambda}$) ratio (logarithmic) at 5.8 and 3.6 $\mu$ versus flux ratio at 8 and 4.5 $\mu$. Small points: SERVS+SWIRE 18436 sources; Triangles (enclosed in yellow edges): 115 sources in common with the selected sample (see Sect. 5.3.2). Cyan stars: stars; Diamonds: SN. Crosses: X-ray detected sources. The colours of the triangles and of the small points from red (extended sources) to blue (pointlike sources) indicate the increasing stellarity (as in Fig. 5). The solid line is the \cite{lacy2004} region and the dashed line the \cite{donley2012} region.
              }
         \label{lacy}
   \end{figure}

 Figure \ref{lacy} shows the diagnostic developed by \cite{lacy2004,lacy2007}, where the stellarity index is coded from red to blue to represent sources from extended to point-like (as in Fig. 5). %% refers to the measurement done in the $r$ band reported in SERVS+SWIRE catalogue (for the sources in SERVS+SWIRE) and to the VST catalogue for the variable sources.
Fig. \ref{lacy} shows sources in the SERVS+SWIRE catalogues and those of the VST-selected sample with available 3.6, 4.5, 5.8, and 8.0 $\mu $ photometry (115 sources).
Owing to the different dust content and temperature, normal galaxies, star-forming galaxies and AGNs occupy different regions of this diagram. This allows, as shown 
in \cite{lacy2004}, defining an empirical wedge that encloses a large portion of the AGN population (as mentioned in Sect. \ref{introduction}).
%Star-forming galaxies are characterised by high PAH signatures at 6.2 $\mu$ and 8 $\mu$, while AGN \textbf{have powerlaw SED due to hot dust emission.}

\cite{donley2012} simulated normal and star-forming galaxies with different levels of AGN contribution to their total emission, in order to derive their IR colours. They find that all the
galaxies with a $>$80\% contribution (between 1 and 10 $\mu$) from an active nucleus are enclosed within the \cite{lacy2004} wedge. Below this limit some AGNs will be missed by the \cite{lacy2004} criterion, to the point where only a few sources with an AGN contribution of $<$20\% can be found inside the wedge. On the other hand, analysing the correspondence between X-ray and IR selection, they found that the Lacy region also contains a significant fraction of non-X-ray detected sources, some of which are likely starbursts. This result is also confirmed by \cite{yun2008}, who found that submillimetre galaxies (which are dominated by starbursts) contaminate the \cite{lacy2004} region. To improve the purity of the IR-selected AGN sample, \cite{donley2012} defined a more restrictive criterion for reducing the starburst contamination to IR-selected AGN samples, which is shown in Figure \ref{lacy}.

Out of the 115 sources of the selected sample represented in the plot, 103 lie within the \cite{lacy2004} wedge, marking them as likely AGN. Although starburst galaxies contaminate this region, their variability supports the AGN classification. 

However, false-positive variability detections are still possible.
Since the Lacy wedge is contaminated by starbursts, we expect that some of the false positive objects may not be AGN even if
  they lie in this region. To estimate the number of such spurious confirmations, we need to evaluate the number of false-positive variable sources that are at the same time contaminants lying within the Lacy wedge. 
 The contamination rate of the Lacy wedge is
  strongly dependent on the flux limits of the survey, as shown in
  \cite{donley2012}. In surveys with fluxes of $>$50 $\mu$Jy at 5.8 $\mu m$, the
  contamination in the majority of the Lacy area is of the order of $ 8$\% (see
  \citealt{lacy2007}).  Decreasing the
  flux limits down to $\sim$11$\mu$Jy, the contamination increases especially at bluer colors (i.e. towards the bottom left edge of the wedge, see Fig. 8 of \citealt{donley2012}). The 5.6 $\mu m$ flux limit
  of our master catalogue
  (with $r<$ 23 mag) is $\sim$10$\mu$Jy, so we expect potential contamination up to 90\% at the leftmost edge of the Lacy region but less than 10 \% \footnote{15\% of the X-ray undetected sources -which in turn are 50\% of the total sample- have colors consistent with star formation as discussed in Fig. 8 and Sect. 9.1 of \citealt{donley2012}} within the Donley wedge. 
 Since our variable sources tend to lie far from the Lacy edges (as can be seen in Fig. 6), we can assume for our estimates an average contamination of 50\%. 

%% On the other hand, the fraction of spurious variable sources within the Lacy wedge can be evaluated considering that in a random subsample of our VST sources the expected number of false positives is <1\%
%% \footnote{estimated from 235/33294$\sim$0.7\%$<$1\% where 235 are the sources above the 3 r.m.s. threshold while 18954+14340=33294 are the total number (see Sect. 3 and 4)}, estimated assuming the worst case scenario where all our sources above the variability threshold are false detections. 

The percentage of spurious variable sources within the Lacy wedge can be
estimated assuming the worst case scenario in which all our sources
above the variability threshold are false positives (1\% of parent sample).
Since the number of sources within the the Lacy wedge is $\sim$2000, we estimate at most 20 false variable objects. 
Of these only 50\% are expected to be also contaminants with colours consistent with AGNs (as discussed above), so we estimate less than ten spurious sources and an upper limit of 10\% (10/103) on the fraction of erroneously confirmed AGN.

We also note from Fig. 6 that the average stellarity
  index of the variable candidates inside the \cite{lacy2004} region decreases
  towards the left-hand side of the diagram; i.e., the sources become more extended, indicating that they are likely to be low redshift galaxies where the nucleus does not dominate the overall emission.
 According to \cite{trevese2008}, many `variable galaxies', i.e. extended variable sources, are narrow emission line galaxies with a low ionisation narrow emission region (LINER).
The majority of point-like sources lie within the \cite{donley2012} wedge, strengthening the view that the \cite{donley2012} region is occupied prevalently by AGN-dominated galaxies. 
Indeed the Donley criterion allows rather complete ($\sim$ 88\% with respect to X-ray selection) samples of luminous AGNs to be identified but misses low-luminosity AGNs with host-dominated SEDs (see \cite{donley2012} for details).
Moreover, all the X-ray detected sources (discussed in Sect. 5.2) are found inside the Donley wedge, further supporting the view that their X-ray emission is not due to star formation but to the active nucleus. These results suggest a strong correspondence between variability, X-ray, and IR selection criterion, at least for AGN-dominated galaxies.

Out of the 12 sources located outside the AGN area of \cite{lacy2004}, three are point-like sources classified as stars in the plot $r-3.6$ versus $r-i$ discussed in Sect. 5.3.1 (ID: 26458, 3121, and 993), and two (IDs 90077 and 48377) have been identified as SN in Sect. 5.1, and are thus likely star-forming galaxies hosting SNe.
The remaining seven sources are difficult to classify with this diagnostic: four pointlike sources (IDs 65830, 1158, 26548, 64257 in Table 3) are located in the ‘blue clump’ in the lower left of the diagram; because of their non-AGN colours and pointlike profiles they can be unresolved galaxies or stars, see \cite{donley2012}.  The remaining three outliers are extended sources: source 109802 (very extended source, with strong variability significance, see Table 3), source 86211 (it has a nearby companion and for this reason it has been flagged with quality flag 2), and source 69697. (This source has quality flag 2 and displays only marginal variability.)
The location of these three extended sources in the diagnostic plot is similar to that of simulated sources with redshift between $\sim$1 and $\sim$2 and AGN contribution below 40\% \citep{donley2012}.

Summarising, according to the IR diagnostic discussed in this section, our sample contains 103/115 likely AGN, with at most ten contaminants.

%______________________________________________________________

\section{Summary and conclusions}\label{conclusions}
In this paper we used the VST SUDARE-VOICE multi-epoch observations in the CDFS area of 2 $\rm deg^2$ to select a sample of AGN via their optical variability. Variability is a valuable tool for selecting AGN candidates, as shown in \cite{sarajedini2003,sarajedini2006,sarajedini2011}, for example.

\cite{trevese2008} used ground-based observations over nine epochs of the ESO/MPI telescope in the framework of the STRESS survey. The variability study conducted in that context in the V band produced results consistent with those obtained in independent X-ray and IR variability-selection methods. They find that the variability selection produces a sample with 44 \% completeness (with respect to the X-ray selected sample with $L_X>10^{42}erg/s$ and known spectra, see \cite{trevese2008} for details) and  less than 40 \% contamination.
The temporal sampling and the number of epochs used in that work were not high enough to limit the contamination and assure a higher level of completeness.
 By increasing the number of epochs, the temporal sampling, and the photometric accuracy, variability surveys can reach far higher completeness: for example, $\sim$ 90 \% completeness was reached in \cite{sesar2007} -with respect to the sample of spectroscopically confirmed SDSS quasars from Stripe 82- exploiting observations spanning a period of six years (one or two observations per season).

In this paper, we have presented a new sample of AGN selected through variability and we have assessed its purity on the basis of IR colors, optical spectroscopic and X-ray surveys over the explored region. The data used here are the VST observations of the CDFS region acquired by the SUDARE-VOICE collaboration. 
We measured the lightcurves of all sources over the four-to-five- month baseline covered by the VST observations. Our selection method relied on the comparison of the rms variability of each source to the average of the entire population of sources with comparable magnitude, selecting as variable those sources with an rms larger than three standard deviations from the average r.m.s.

 From the catalogues of sources detected in more than six epochs and $r$ $<$ 23 mag, we identified 175 candidates (we refer to them as `selected sample'). 
To validate the sample, we used information available both within the VST-SUDARE consortium and in the literature (\cite{hsu2014}, \cite{boutsia2009}, \cite{lonsdale2004}, \citealt{mauduit2012}). 

  \begin{table}
\centering
\begin{tabular}{c c c}        
 & & N \\
%(1) & Detected & 80896 \\
%(2) & Variable & 1013 \\
\hline\hline
(1) & Selected & 175 \\
\hline
(2) & SN & 21/175 \\
(3) & X-rays & 12/15 \\
(4) & AGN (SED) & 12/15 \\
%(6) & AGN-SED & 12\\
%(7) & SERVS-SWIRE & 158 \\
%(8) & g r u & 37 \\
%(9) & g r i & 56 \\
(5) & Stars (r, i, 3.6$\mu$m) & 6/57 \\
(6) & AGN (IR) & 103/115 \\
(7) & AGN (spectra) & 9/9 \\
\hline
%(10) & no-SERVS-SWIRE & 17 \\
%% (13) & Borders & 3 \\
%% (14) & SN & 7 \\
%% (15) & Faint AGN  & 5 \\
%(11) & Bright AGN & 2 \\
\hline
%(17) & z & 21 \\

\end{tabular}\caption{Number of sources of: %(1) the detected sources in the CDFS (in more than 6 epochs, masked), (2) variable sources (3$\sigma$), 
(1) selected sample, (2)  SN candidates (see Sect. 5.1), (3)  Candidates with X-ray detection (\cite{hsu2014}, see Sect. 5.2), (4) AGN candidates (AGN SED in \cite{hsu2014}, see Sect. 5.2), (5) Star candidates (diagnostic $r-i$ versus 3.6 $\mu$m to $r$ band flux ratios, see Sect. 5.3.1), (6) AGN \cite{lacy2004} diagnostic using IR colours, see Sect. 5.3.2), (7) AGN candidates confirmed in optical spectroscopy by \cite{boutsia2009} (BLAGN, Sect. 5.2)}\label{tsummary}
\end{table}

We found that:
\begin{enumerate}
\item 12$\pm$2\% (21/175) of the selected sample are classified as SN, based on both visual inspection of the lightcurves and template fitting by the SUDARE-I collaboration (Cappellaro et al., in preparation).
\item An additional 3\% (6/175) of the selected sample includes stars, based on the optical-IR diagnostic ($r-i$ versus the 3.6 $\mu$m to $r$ band flux ratio) of \cite{rowan-robinson2013}. This diagram also shows that most of the remaining variable sources have colours consistent with AGN
\item For the 115 objects with detections in all four IRAC bands, 103 (59\% of the selected sample) have colours that are consistent with AGN emission. Although the \cite{lacy2004} criterion allows for contamination by starforming galaxies, the detection of variability supports the idea that in these sources a significant fraction of the IR emission is produced by an AGN. We estimate an upper limit of 10\% to the number of erroneously confirmed AGNs. We found that the most compact sources tend to have colours consistent with the more restrictive criterion proposed by \cite{donley2012}, which are AGN-dominated objects, while the most extended sources lie on the IR diagram where some contamination from starburst/host galaxy light is expected. Outside the wedge of \cite{lacy2004} we identified twelve sources: five of these are among the SNe and stars discussed above, and the remaining seven are objects without clear AGN/star/SN signatures (except possibly for the variability properties) and may be spurious detections or normal galaxies hosting very low luminosity AGN.
\item In the selected sample there are 15 sources within the ECDFS area for which we have SED, X-ray, and spectroscopic information presented in \citep{hsu2014} and \cite{boutsia2009}.  In particular we found:
\begin{itemize}
\item twelve sources with a clear AGN signature in their SED, which are also X-ray detected. All of them are AGN, based on the IR selection (see Table \ref{tresults});
\item three variable sources with purely galactic (non-AGN) SED and not detected in the X-ray band; this is consistent with our previous SN classification based on lightcurves;
\item 
nine variable sources confirmed by the optical spectroscopy presented in \cite{boutsia2009}, which have been classified as BLAGN. All of them are already classified as AGN based on the IR selection, X-ray selection, and SED properties. This result is consistent with the one shown in Paper I: the majority of the variable sources in COSMOS are BLAGN (as expected given the selection method).
\end{itemize}
\end{enumerate}

Overall, the total number of candidates for which we could employ the diagnostics discussed in Sect. 5 (as can be seen in Table 3) is 137 out of a total 175 selected candidates. Of the 137 sources, we found 102 AGN confirmed by at least one diagnostic, which is 58\% of the selected sample. Among the remaining sources six are stars, and nine are SN. Column 13 of Table 3 indicates the sum of the AGN flags in Cols. 9, 10, 11, 12. 
For AGN, this is a positive number that counts the number of positive diagnostics.
 The flag is instead negative for stars and SN. The percentage of confirmed AGN, as well as those of identified contaminants are consistent with the results presented in Paper I for the COSMOS region. We also found 18 objects whose AGN nature could not be verified by the diagnostics investigated in this work. There is, finally, a group of 38 sources (among which 12 SN) for which we could not employ any diagnostic other than our lightcurve inspection.

Our results are useful in predicting the expected performance of future monitoring surveys similar in depth and cadence to our own, to assemble samples of AGNs based on variability alone. Twelve percent of the selected sample is composed of SNe (identified through the inspection of the lightcurves), which were removed from the sample. In our dataset, 4\% (6/154) of the non-SN sources are stars, while 66\% (102/154) are confirmed AGNs based on the available diagnostics. We can thus estimate an upper limit to the contamination of the variability-selected AGN sample of $\simeq$ 34 \%, assuming a worst-case scenario where all unconfirmed variable sources (excluding SN) are non-AGNs. However we point out that restricting the analysis to the sources with available multi-wavelength ancillary information, the purity of our sample is  $\sim 80$ \% (102 confirmed AGNs/128 non-SN). While some of the IR confirmed AGNs can be spurious (as discussed in Sect. 5.3.2) we estimate that this effect would yield a false positive rate of at most 10\%. Our results agree with those found in Paper I on the COSMOS field where almost all sources could be identified by means of multi-wavelength information and/or spectroscopy; in that case, we showed that our variability-selected sample reached a final purity of 93\% (Paper I).
%% As discussed in Sec. 5.3.2, since the IR-selected AGN sample based on the Donley criterion is not free from contamination itself, some of the IR selected sources could be starforming galaxies. However, based on the comparison with X-ray selected samples, X-ray stacking, and color-selected populations (BzK: \cite{daddi2004}; DRGs: e.g. \cite{franx2003,vandokkum2003,conselice2007,lane2007}; LBGs: e.g., \citealt{steidel2003,shim2011}), Donley and collaborators estimate that this contamination is $<10$\% and possibly as low as a few percent given the multi-wavelength photometric and spectral properties of the unconfirmed candidates, which favour their AGN nature. 
 A spectroscopic follow-up of the unconfirmed sources is desirable to conclusively verify this result for the CDFS area.

The completeness of the variability-selected survey presented in this work is 22 \%, computed with respect to the IR selection of \cite{donley2012} described in Sect. 5.3.2.
In Paper I, the completeness (computed with respect to X-ray samples) has been estimated to be 15 \% for a five-month baseline; the two results are thus in broad agreement considering that the completeness is estimated with respect to different reference populations (for instance, only 50\% of the IR selected sources by Donley are detected in X-rays, but at the same time the IR criterion tends to favour luminous AGNs with $L_X>10^{43}$ erg s$^{-1}$).
The completeness found in the present work and in Paper I represents a lower limit to the potential completeness of the variability-selected survey that improves extending the monitoring baseline, as we have seen in previous papers (e.g. \cite {sesar2007} and Paper 1).
The new observations that are currently being acquired for the COSMOS field (P.I: G. Pignata) with VST will allow us to directly compare these results using a three-year long monitoring baseline.

Summarising, this paper makes use of the VST observations, together with the IR photometry, SED fitting and X-ray information where available to confirm the nature of the AGN candidates. The IR data, available over the full survey area, allow confirming the consistency of the variability selection with the IR colour selection method, while the detection of variability may prove useful to detect the presence of an AGN in IR-selected starburst galaxies.

  By applying the variability-selection method used here to future datasets such as LSST, it will be possible to complete the AGN demography including strongly variable sources and also AGN, which could be confused with stars of similar colours. 

 \tiny{
\onllongtab{
\begin{longtable}{l c c c c c c c c c c c c}
\caption{Results on the selected sample.
(1) optical ID; (2) coordinates (J2000); (3) average $r$ band magnitude; (4) stellarity index (from Sextractor, see text); (5) significance of variability in r.m.s. units; (6) quality flag defined in Sect. 4; (7) SN (identified by their LC and confirmed in SUDARE-I, see Sect. 5.1); (8) stars identified with the diagnostic using $r-i$ versus the 3.6 $\mu m$ to $r$ band flux ratio (see Sect. 5.3.1);
 (9) X-ray detected/ non-detected sources flagged with 1 and 0, respectively (see Sect. 5.2); (10) AGN validation by optical spectroscopy \citep{boutsia2009}; (11) sources matching the \cite{lacy2004} IR criterion for AGN selection (see Sect. 5.3.2); (12) AGN validation by multi-wavelength SED fitting \citep{rowan-robinson2013,hsu2014}: sources fitted by an AGN template are flagged with 1, otherwise with 0 (slant characters highlight sources in \cite{rowan-robinson2013}, regular ones the sources in \cite{hsu2014} (13) Source classification. This flag is obtained summing the individual AGN flags reported in Cols. (9), (10), (11), (12). It is, thus, a positive number for AGN. The SN are indicated with flag -2 and the stars with flag 1.}\label{tresults} \\
\hline
\hline

ID  & RA,DEC  & $<r>$ & Stellarity & $\sigma^{*}$ & Quality flag & SN& Star  & AGN  & AGN & AGN  & AGN & Classification \\
   & deg & mag &  & &  & LC  & r,i, 3.6$\mu m$ & X-ray & spectra & IR & SED & tot \\
(1) & (2) & (3) & (4) & (5) & (6) & (7) & (8) & (9) & (10) & (11) & (12) & (13) %  \footnote{Columns: ....}
  \\ 
\endfirsthead
\hline
\endhead

\hline
\hline
     6384   &    03:27:55.86  -28:00:06.5  &   21.59   &   0.98   &  4.5   & 1 &   0   &   1   &  -   &  -   &   -   &  -   &   -1   \\
   6690   &    03:28:30.02  -27:59:55.7  &   22.7   &   0.88   &  8.3   & 1 &   1   &   0   &  -   &  -   &   -   &  \emph{0}   &   -2   \\
   13372   &    03:27:12.54  -27:55:09.5  &   21.67   &   0.03   &  3.3   & 1 &   0   &   -   &  -   &  -   &   -   &  -   &   -   \\
   16666   &    03:29:23.05  -27:52:59.4  &   22.32   &   0.04   &  7.4   & 1 &   1   &   -   &  -   &  -   &   -   &  -   &   -2   \\
   20470   &    03:27:49.38  -27:50:15.0  &   22.64   &   0.96   &  39.0   & 1 &   1   &   -   &  -   &  -   &   -   &  -   &   -2   \\
   24366   &    03:27:53.83  -27:47:21.5  &   21.01   &   0.97   &  3.2   & 1 &   0   &   -   &  -   &  -   &   -   &  -   &   -   \\
   27357   &    03:29:21.29  -27:45:09.1  &   22.26   &   0.84   &  4.2   & 1 &   0   &   -   &  -   &  -   &   -   &  -   &   -   \\
   29449   &    03:29:02.74  -27:43:48.0  &   22.49   &   0.11   &  11.0   & 1 &   1   &   -   &  -   &  -   &   -   &  -   &   -2   \\
   30173   &    03:29:55.41  -27:43:18.3  &   22.82   &   0.97   &  4.4   & 1 &   0   &   0   &  -   &  -   &   -   &  \emph{0}   &   0   \\
   34660   &    03:30:19.49  -27:40:30.4  &   21.98   &   0.97   &  5.4   & 1 &   0   &   0   &  -   &  -   &   -   &  \emph{0}   &   0   \\
   41682   &    03:29:38.52  -27:36:19.0  &   21.94   &   0.9   &  7.2   & 1 &   0   &   0   &  -   &  -   &   -   &  \emph{0}   &   0   \\
   44289   &    03:30:32.83  -27:34:46.9  &   22.99   &   0.89   &  15.8   & 1 &   0   &   -   &  -   &  -   &   -   &  -   &   -   \\
   47325   &    03:27:59.79  -27:32:55.4  &   21.62   &   0.94   &  3.4   & 1 &   0   &   0   &  -   &  -   &   -   &  \emph{0}   &   0   \\
   49662   &    03:30:26.11  -27:31:22.2  &   21.56   &   0.03   &  7.1   & 1 &   1   &   -   &  -   &  -   &   -   &  \emph{0}   &   -2   \\
   54759   &    03:30:27.32  -27:28:13.2  &   22.78   &   0.01   &  12.0   & 1 &   1   &   -   &  -   &  -   &   -   &  -   &   -2   \\
   66439   &    03:31:04.81  -27:20:32.6  &   22.77   &   0.31   &  12.7   & 1 &   1   &   -   &  -   &  -   &   -   &  -   &   -2   \\
   75327   &    03:27:42.78  -27:14:45.4  &   22.75   &   0.98   &  4.2   & 1 &   0   &   -   &  -   &  -   &   -   &  -   &   -   \\
   79422   &    03:28:49.24  -27:12:29.4  &   22.88   &   0.96   &  5.5   & 1 &   0   &   -   &  -   &  -   &   -   &  -   &   -   \\
   2431   &    03:35:32.96  -28:03:24.9  &   22.12   &   0.98   &  3.9   & 1 &   0   &   0   &  -   &  -   &   -   &  -   &   -   \\
   2435   &    03:35:15.05  -28:03:25.5  &   22.39   &   0.78   &  3.3   & 1 &   0   &   0   &  -   &  -   &   -   &  \emph{0}   &   0   \\
   5854   &    03:32:15.43  -28:01:23.2  &   22.47   &   0.03   &  8.4   & 1 &   1   &   0   &  0   &  -   &   -   &  0   &   -2   \\
   7120   &    03:34:49.59  -28:00:49.7  &   21.0   &   0.03   &  6.8   & 1 &   1   &   0   &  -   &  -   &   -   &  \emph{0}   &   -2   \\
   15811   &    03:32:08.83  -27:55:44.9  &   20.82   &   0.03   &  6.2   & 1 &   1   &   0   &  0   &  -   &   -   &  0   &   -2   \\
   19508   &    03:34:56.87  -27:53:33.0  &   21.12   &   0.03   &  15.5   & 1 &   1   &   0   &  -   &  -   &   -   &  \emph{0}   &   -2   \\
   29957   &    03:33:57.87  -27:46:46.8  &   22.28   &   0.8   &  3.2   & 1 &   0   &   -   &  -   &  -   &   -   &  -   &   -   \\
   60469   &    03:35:16.28  -27:29:49.1  &   21.05   &   0.03   &  5.1   & 1 &   0   &   -   &  -   &  -   &   -   &  -   &   -   \\
   75131   &    03:33:58.45  -27:22:10.3  &   21.69   &   0.12   &  6.2   & 1 &   1   &   -   &  -   &  -   &   -   &  -   &    -2  \\
   75846   &    03:34:47.93  -27:21:42.6  &   21.51   &   0.04   &  10.0   & 1 &   1   &   -   &  -   &  -   &   -   &  -   &   -2   \\
   78144   &    03:34:38.15  -27:20:20.6  &   22.79   &   0.09   &  11.5   & 1 &   1   &   -   &  -   &  -   &   -   &  -   &   -2   \\
   85754   &    03:33:43.87  -27:15:56.7  &   22.1   &   0.8   &  4.1   & 1 &   0   &   -   &  -   &  -   &   -   &  -   &   -   \\
   99982   &    03:33:26.81  -27:08:50.5  &   21.34   &   0.03   &  10.9   & 1 &   1   &   -   &  -   &  -   &   -   &  -   &   -2   \\
   100845   &    03:33:03.60  -27:08:25.4  &   21.55   &   0.93   &  6.1   & 1 &   0   &   -   &  -   &  -   &   -   &  -   &   -   \\
   105093   &    03:35:02.03  -27:06:16.0  &   22.48   &   0.81   &  12.2   & 1 &   0   &   -   &  -   &  -   &   -   &  -   &   -   \\
   2743   &    03:29:59.44  -28:02:18.0  &   22.2   &   0.47   &  4.5   & 1 &   0   &   -   &  -   &  -   &   -   &  -   &   -   \\
   28903   &    03:30:07.49  -27:44:09.1  &   22.78   &   0.93   &  10.8   & 1 &   0   &   0   &  -   &  -   &   -   &  \emph{0}   &   0   \\
   32669   &    03:30:43.25  -27:41:41.3  &   21.52   &   0.94   &  3.1   & 1 &   0   &   0   &  -   &  -   &   -   &  \emph{0}   &   0   \\
   36393   &    03:30:12.85  -27:39:31.2  &   22.93   &   0.94   &  3.4   & 1 &   0   &   0   &  -   &  -   &   -   &  \emph{0}   &   0   \\
   47852   &    03:27:19.66  -27:32:30.4  &   20.12   &   0.9   &  4.2   & 1 &   0   &   -   &  -   &  -   &   -   &  -   &   -   \\
   67668   &    03:30:06.41  -27:19:48.3  &   22.32   &   0.02   &  4.1   & 1 &   1   &   -   &  -   &  -   &   -   &  -   &   -2   \\
   69628   &    03:28:31.63  -27:18:34.2  &   22.11   &   0.03   &  4.6   & 1 &   0   &   -   &  -   &  -   &   -   &  -   &   -   \\
   70019   &    03:30:28.14  -27:18:17.5  &   22.88   &   0.04   &  3.9   & 1 &   0   &   -   &  -   &  -   &   -   &  -   &   -   \\
   70305   &    03:29:40.46  -27:18:06.7  &   20.82   &   0.03   &  3.4   & 1 &   0   &   -   &  -   &  -   &   -   &  -   &   -   \\
   76670   &    03:30:25.09  -27:13:57.7  &   21.41   &   0.89   &  3.7   & 1 &   0   &   -   &  -   &  -   &   -   &  -   &   -   \\
   89177   &    03:28:00.28  -27:07:14.1  &   18.52   &   0.87   &  16.6   & 1 &   0   &   -   &  -   &  -   &   -   &  -   &   -   \\
   290   &    03:33:08.76  -28:05:04.1  &   22.8   &   0.46   &  25.6   & 1 &   1   &   -   &  -   &  -   &   -   &  -   &   -2   \\
   525   &    03:35:14.36  -28:05:07.4  &   17.08   &   0.96   &  3.6   & 1 &   0   &   1   &  -   &  -   &   -   &  -   &   -1   \\
   618   &    03:31:50.76  -28:04:43.3  &   22.28   &   0.98   &  4.3   & 1 &   0   &   0   &  -   &  -   &   -   &  -   &   0   \\
   749   &    03:35:17.41  -28:04:39.1  &   18.21   &   0.89   &  3.2   & 1 &   0   &   1   &  -   &  -   &   -   &  -   &   -1   \\
   11324   &    03:32:15.84  -27:58:22.7  &   22.81   &   0.04   &  3.5   & 1 &   1   &   0   &  0   &  -   &   -   &  0   &   -2   \\
   72386   &    03:31:29.07  -27:23:36.5  &   21.1   &   0.88   &  3.0   & 1 &   0   &   -   &  -   &  -   &   -   &  -   &   -   \\
   82256   &    03:35:00.42  -27:17:58.3  &   21.04   &   0.03   &  5.3   & 1 &   0   &   -   &  -   &  -   &   -   &  -   &   -   \\
   82948   &    03:33:39.34  -27:17:31.3  &   21.87   &   0.98   &  3.4   & 1 &   0   &   -   &  -   &  -   &   -   &  -   &   -   \\
   84628   &    03:33:57.54  -27:16:32.3  &   22.73   &   0.97   &  3.0   & 1 &   0   &   -   &  -   &  -   &   -   &  -   &   -   \\
   86007   &    03:35:17.02  -27:15:46.8  &   22.72   &   0.98   &  11.3   & 1 &   0   &   -   &  -   &  -   &   -   &  -   &   -   \\
   94094   &    03:35:15.03  -27:11:44.8  &   22.52   &   0.98   &  6.4   & 1 &   0   &   -   &  -   &  -   &   -   &  -   &   -   \\
   108618   &    03:34:10.32  -27:04:07.0  &   19.06   &   0.86   &  4.2   & 1 &   0   &   -   &  -   &  -   &   -   &  -   &   -   \\
   110212   &    03:31:56.21  -27:02:56.4  &   20.02   &   0.9   &  4.0   & 1 &   0   &   -   &  -   &  -   &   -   &  -   &   -   \\
   110606   &    03:33:32.72  -27:02:41.8  &   20.62   &   0.03   &  3.4   & 1 &   0   &   -   &  -   &  -   &   -   &  -   &   -   \\
   1537   &    03:28:13.01  -28:03:11.5  &   20.84   &   0.92   &  3.6   & 1 &   0   &   0   &  -   &  -   &   1   &  \emph{1}   &   2   \\
   2425   &    03:30:17.80  -28:02:31.3  &   20.33   &   0.88   &  7.9   & 1 &   0   &   0   &  -   &  -   &   1   &  \emph{1}   &   2   \\
   2535   &    03:28:18.32  -28:02:27.7  &   20.31   &   0.89   &  6.8   & 1 &   0   &   0   &  -   &  -   &   1   &  \emph{1}   &   2   \\
   2697   &    03:30:15.97  -28:02:19.8  &   19.93   &   0.86   &  8.8   & 1 &   0   &   0   &  -   &  -   &   1   &  \emph{1}   &   2   \\
   6442   &    03:27:11.01  -28:00:02.9  &   20.8   &   0.04   &  3.6   & 1 &   0   &   -   &  -   &  -   &   1   &  -   &   1   \\
   7134   &    03:30:02.67  -27:59:37.0  &   21.52   &   0.1   &  7.9   & 1 &   0   &   0   &  -   &  -   &   1   &  \emph{1}   &   2   \\
   8056   &    03:30:05.03  -27:58:55.2  &   21.89   &   0.07   &  4.9   & 1 &   0   &   0   &  -   &  -   &   1   &  \emph{0}   &   1   \\
   12153   &    03:28:56.48  -27:56:00.9  &   21.61   &   0.92   &  9.8   & 1 &   0   &   0   &  -   &  -   &   1   &  \emph{0}   &   1   \\
   16716   &    03:29:24.24  -27:52:56.9  &   21.62   &   0.92   &  4.7   & 1 &   0   &   0   &  -   &  -   &   1   &  \emph{0}   &   1   \\
   18350   &    03:29:40.47  -27:51:43.6  &   19.55   &   0.87   &  5.9   & 1 &   0   &   0   &  -   &  -   &   1   &  \emph{1}   &   2   \\
   21592   &    03:30:52.20  -27:49:26.7  &   21.12   &   0.94   &  6.3   & 1 &   0   &   -   &  -   &  -   &   1   &  -   &   1   \\
   27638   &    03:30:39.70  -27:44:55.4  &   21.92   &   0.87   &  11.6   & 1 &   0   &   -   &  -   &  -   &   1   &  -   &   1   \\
   28499   &    03:27:03.63  -27:44:25.2  &   19.05   &   1.0   &  6.4   & 1 &   0   &   -   &  -   &  -   &   1   &  -   &   1   \\
   29899   &    03:27:52.98  -27:43:28.7  &   21.28   &   0.92   &  5.1   & 1 &   0   &   -   &  -   &  -   &   0   &  -   &   0   \\
   32210   &    03:27:24.94  -27:42:02.6  &   19.55   &   0.92   &  8.1   & 1 &   0   &   -   &  -   &  -   &   1   &  -   &   1   \\
   33242   &    03:29:56.70  -27:41:22.7  &   20.96   &   0.85   &  9.8   & 1 &   1   &   0   &  -   &  -   &   1   &  \emph{0}   &   -2   \\
   34969   &    03:29:39.32  -27:40:20.2  &   21.74   &   0.92   &  3.9   & 1 &   0   &   0   &  -   &  -   &   1   &  \emph{0}   &   1   \\
   35974   &    03:29:02.09  -27:39:46.7  &   21.39   &   0.96   &  7.2   & 1 &   0   &   0   &  -   &  -   &   1   &  \emph{0}   &   1   \\
   39556   &    03:30:28.11  -27:37:36.6  &   19.87   &   0.87   &  4.6   & 1 &   0   &   0   &  -   &  -   &   1   &  \emph{1}   &   2   \\
   45234   &    03:28:37.76  -27:34:15.4  &   20.47   &   0.92   &  4.4   & 1 &   0   &   0   &  -   &  -   &   1   &  \emph{0}   &   1   \\
   55181   &    03:28:46.21  -27:27:58.5  &   21.89   &   0.85   &  12.0   & 1 &   0   &   -   &  -   &  -   &   1   &  -   &   1   \\
   59923   &    03:28:51.64  -27:24:53.4  &   22.54   &   0.54   &  4.8   & 1 &   0   &   -   &  -   &  -   &   1   &  -   &   1   \\
   63050   &    03:30:05.75  -27:22:48.6  &   21.62   &   0.93   &  9.6   & 1 &   0   &   -   &  -   &  -   &   1   &  -   &   1   \\
   65830   &    03:30:14.36  -27:21:01.2  &   20.14   &   0.89   &  4.6   & 1 &   0   &   -   &  -   &  -   &   0   &  -   &   0   \\
   66781   &    03:30:05.56  -27:20:20.6  &   21.85   &   0.95   &  3.0   & 1 &   0   &   -   &  -   &  -   &   1   &  -   &   1   \\
   69322   &    03:27:36.64  -27:18:42.7  &   20.72   &   0.91   &  8.0   & 1 &   0   &   -   &  -   &  -   &   1   &  -   &   1   \\
   69325   &    03:28:38.02  -27:18:44.0  &   20.87   &   0.91   &  17.2   & 1 &   0   &   -   &  -   &  -   &   1   &  -   &   1   \\
   72080   &    03:27:21.61  -27:16:49.3  &   22.49   &   0.13   &  4.8   & 1 &   0   &   -   &  -   &  -   &   1   &  -   &   1   \\
   76439   &    03:29:31.00  -27:14:07.8  &   21.84   &   0.94   &  6.0   & 1 &   0   &   -   &  -   &  -   &   1   &  -   &   1   \\
   79350   &    03:27:55.68  -27:12:30.9  &   22.0   &   0.98   &  4.4   & 1 &   0   &   -   &  -   &  -   &   1   &  -   &   1   \\
   80235   &    03:28:50.23  -27:12:08.1  &   19.17   &   0.9   &  16.5   & 1 &   0   &   -   &  -   &  -   &   1   &  -   &   1   \\
   82885   &    03:29:25.35  -27:10:52.5  &   19.45   &   0.89   &  4.4   & 1 &   0   &   -   &  -   &  -   &   1   &  -   &   1   \\
   85284   &    03:27:24.97  -27:09:20.4  &   20.67   &   0.65   &  9.5   & 1 &   0   &   -   &  -   &  -   &   1   &  -   &   1   \\
   90638   &    03:27:17.13  -27:06:17.9  &   19.6   &   0.85   &  3.3   & 1 &   0   &   -   &  -   &  -   &   1   &  -   &   1   \\
   94735   &    03:30:54.57  -27:03:40.7  &   20.99   &   0.91   &  4.7   & 1 &   0   &   -   &  -   &  -   &   1   &  -   &   1   \\
   95399   &    03:28:55.70  -27:03:15.2  &   21.6   &   0.95   &  9.0   & 1 &   0   &   -   &  -   &  -   &   1   &  -   &   1   \\
   2859   &    03:32:31.99  -28:03:10.1  &   19.5   &   0.85   &  7.4   & 1 &   0   &   0   &  1   &  1   &   1   &  1   &   4   \\
   4362   &    03:32:20.32  -28:02:15.1  &   20.56   &   0.91   &  10.7   & 1 &   0   &   0   &  1   &  1   &   1   &  1   &   4   \\
   6814   &    03:31:27.79  -28:00:51.2  &   21.99   &   0.95   &  3.2   & 1 &   0   &   0   &  1   &  1   &   1   &  1   &   4   \\
   9806   &    03:34:39.03  -27:59:15.4  &   20.54   &   0.03   &  4.0   & 1 &   0   &   -   &  -   &  -   &   1   &  -   &   1   \\
   15209   &    03:34:42.03  -27:56:05.8  &   20.89   &   0.89   &  6.0   & 1 &   0   &   0   &  -   &  -   &   1   &  \emph{0}   &   1   \\
   17415   &    03:31:45.21  -27:54:35.8  &   20.65   &   0.9   &  14.2   & 1 &   0   &   0   &  1   &  -   &   1   &  1   &   3   \\
   26084   &    03:33:32.75  -27:49:08.1  &   21.87   &   0.99   &  4.5   & 1 &   0   &   -   &  1   &  1   &   1   &  -   &   3   \\
   27645   &    03:35:29.27  -27:48:07.6  &   20.64   &   0.91   &  6.3   & 1 &   0   &   -   &  -   &  -   &   1   &  -   &   1   \\
   28275   &    03:32:59.83  -27:47:48.4  &   20.9   &   0.93   &  3.9   & 1 &   0   &   -   &  1   &  1   &   1   &  -   &   3   \\
   28432   &    03:34:52.50  -27:47:41.1  &   21.68   &   0.92   &  4.3   & 1 &   0   &   -   &  -   &  -   &   1   &  -   &   1   \\
   35698   &    03:34:08.26  -27:43:38.0  &   20.21   &   0.89   &  3.6   & 1 &   0   &   -   &  -   &  -   &   1   &  -   &   1   \\
   41079   &    03:32:26.49  -27:40:35.7  &   19.9   &   0.89   &  4.1   & 1 &   0   &   0   &  1   &  1   &   1   &  1   &   4   \\
   42395   &    03:34:42.43  -27:39:51.6  &   20.89   &   0.94   &  12.0   & 1 &   0   &   0   &  -   &  -   &   1   &  \emph{1}   &   2   \\
   42982   &    03:32:16.19  -27:39:30.4  &   20.12   &   0.88   &  6.1   & 1 &   0   &   0   &  1   &  1   &   1   &  1   &   4   \\
   46587   &    03:32:11.64  -27:37:26.0  &   18.78   &   0.86   &  5.1   & 1 &   0   &   0   &  1   &  -   &   1   &  1   &   3   \\
   47327   &    03:35:24.94  -27:36:55.6  &   22.02   &   0.91   &  3.4   & 1 &   0   &   0   &  -   &  -   &   1   &  \emph{0}   &   1   \\
   48377   &    03:35:28.30  -27:36:21.5  &   20.26   &   0.03   &  7.3   & 1 &   1   &   0   &  -   &  -   &   0   &  \emph{0}   &   -2   \\
   60497   &    03:34:59.84  -27:29:56.9  &   19.81   &   0.03   &  4.7   & 1 &   0   &   -   &  -   &  -   &   1   &  -   &   1   \\
   62569   &    03:33:37.75  -27:28:46.3  &   20.32   &   0.07   &  3.9   & 1 &   0   &   -   &  -   &  -   &   1   &  -   &   1   \\
   63435   &    03:32:55.70  -27:28:17.3  &   22.4   &   0.9   &  4.4   & 1 &   0   &   -   &  -   &  -   &   1   &  -   &   1   \\
   64886   &    03:31:56.25  -27:27:30.7  &   21.59   &   0.93   &  3.8   & 1 &   0   &   -   &  -   &  -   &   1   &  -   &   1   \\
   66489   &    03:35:24.18  -27:26:39.6  &   22.44   &   0.98   &  5.2   & 1 &   0   &   -   &  -   &  -   &   1   &  -   &   1   \\
   68199   &    03:34:21.38  -27:25:48.4  &   21.86   &   0.82   &  3.8   & 1 &   0   &   -   &  -   &  -   &   1   &  -   &   1   \\
   71431   &    03:33:31.69  -27:24:09.0  &   21.74   &   0.88   &  10.0   & 1 &   0   &   -   &  -   &  -   &   1   &  -   &   1   \\
   79604   &    03:34:41.51  -27:19:29.4  &   21.38   &   0.96   &  9.0   & 1 &   0   &   -   &  -   &  -   &   1   &  -   &   1   \\
   79779   &    03:34:06.82  -27:19:23.0  &   21.83   &   0.93   &  5.1   & 1 &   0   &   -   &  -   &  -   &   1   &  -   &   1   \\
   81470   &    03:35:31.94  -27:18:25.2  &   18.81   &   0.95   &  8.0   & 1 &   0   &   -   &  -   &  -   &   1   &  -   &   1   \\
   83511   &    03:34:13.21  -27:17:11.9  &   21.12   &   0.93   &  7.7   & 1 &   0   &   -   &  -   &  -   &   1   &  -   &   1   \\
   84399   &    03:32:11.40  -27:16:39.6  &   21.17   &   0.93   &  3.8   & 1 &   0   &   -   &  -   &  -   &   1   &  -   &   1   \\
   85224   &    03:32:02.44  -27:16:18.6  &   18.22   &   0.87   &  7.6   & 1 &   0   &   -   &  -   &  -   &   1   &  -   &   1   \\
   87751   &    03:33:08.11  -27:14:52.4  &   21.29   &   0.09   &  5.6   & 1 &   0   &   -   &  -   &  -   &   1   &  -   &   1   \\
   90088   &    03:33:33.96  -27:13:46.2  &   20.68   &   0.91   &  3.3   & 1 &   0   &   -   &  -   &  -   &   1   &  -   &   1   \\
   90252   &    03:35:09.39  -27:13:40.1  &   20.53   &   0.72   &  5.8   & 1 &   0   &   -   &  -   &  -   &   1   &  -   &   1   \\
   92096   &    03:34:27.83  -27:12:45.2  &   20.6   &   0.91   &  3.4   & 1 &   0   &   -   &  -   &  -   &   1   &  -   &   1   \\
   93300   &    03:35:02.74  -27:12:10.3  &   20.33   &   0.8   &  5.1   & 1 &   0   &   -   &  -   &  -   &   1   &  -   &   1   \\
   94220   &    03:31:39.25  -27:11:40.3  &   21.48   &   0.89   &  3.1   & 1 &   0   &   -   &  -   &  -   &   1   &  -   &   1   \\
   94363   &    03:33:39.51  -27:11:39.0  &   20.81   &   0.91   &  5.8   & 1 &   0   &   -   &  -   &  -   &   1   &  -   &   1   \\
   94868   &    03:35:26.87  -27:11:22.1  &   21.75   &   0.94   &  3.7   & 1 &   0   &   -   &  -   &  -   &   1   &  -   &   1   \\
   95338   &    03:35:16.24  -27:11:08.0  &   20.59   &   0.9   &  6.9   & 1 &   0   &   -   &  -   &  -   &   1   &  -   &   1   \\
   96679   &    03:32:47.09  -27:10:36.3  &   18.62   &   0.81   &  17.0   & 1 &   0   &   -   &  -   &  -   &   1   &  -   &   1   \\
   99186   &    03:33:31.92  -27:09:16.4  &   19.7   &   0.93   &  10.4   & 1 &   0   &   -   &  -   &  -   &   1   &  -   &   1   \\
   99735   &    03:32:03.67  -27:08:55.8  &   21.29   &   0.96   &  5.1   & 1 &   0   &   -   &  -   &  -   &   1   &  -   &   1   \\
   100120   &    03:35:29.23  -27:08:43.4  &   21.04   &   0.91   &  8.7   & 1 &   0   &   -   &  -   &  -   &   1   &  -   &   1   \\
   102426   &    03:33:00.69  -27:07:39.8  &   20.53   &   0.71   &  5.3   & 1 &   0   &   -   &  -   &  -   &   1   &  -   &   1   \\
   107237   &    03:32:39.74  -27:04:58.7  &   20.8   &   0.75   &  15.8   & 1 &   0   &   -   &  -   &  -   &   1   &  -   &   1   \\
   107399   &    03:32:16.66  -27:04:52.8  &   21.69   &   0.98   &  4.6   & 1 &   0   &   -   &  -   &  -   &   1   &  -   &   1   \\
   109802   &    03:31:50.59  -27:03:15.5  &   19.16   &   0.03   &  5.1   & 1 &   0   &   -   &  -   &  -   &   0   &  -   &   0   \\
   993   &    03:30:46.15  -28:03:45.6  &   17.51   &   0.97   &  3.8   & 2 &   0   &   1   &  -   &  -   &   0   &  -   &   -1   \\
   1158   &    03:27:20.05  -28:03:31.1  &   18.37   &   0.88   &  3.8   & 2 &   0   &   -   &  -   &  -   &   0   &  -   &   0   \\
   1806   &    03:29:40.20  -28:02:59.2  &   19.94   &   0.9   &  4.1   & 2 &   0   &   0   &  -   &  -   &   1   &  \emph{1}   &   2   \\
   3121   &    03:29:57.33  -28:02:07.7  &   18.61   &   0.89   &  6.8   & 2 &   0   &   1   &  -   &  -   &   0   &  -   &   -1   \\
   17120   &    03:29:41.57  -27:52:37.5  &   21.19   &   0.91   &  3.5   & 2 &   0   &   0   &  -   &  -   &   1   &  \emph{0}   &   1   \\
   22268   &    03:30:02.52  -27:48:58.0  &   20.25   &   0.95   &  5.6   & 2 &   0   &   0   &  -   &  -   &   1   &  \emph{1}   &   2   \\
   26458   &    03:28:51.34  -27:45:52.2  &   18.73   &   0.84   &  3.4   & 2 &   0   &   1   &  -   &  -   &   0   &  \emph{0}   &   -1   \\
   26548   &    03:28:30.68  -27:45:43.2  &   18.76   &   0.87   &  6.0   & 2 &   0   &   -   &  -   &  -   &   0   &  -   &   0   \\
   37366   &    03:30:12.05  -27:38:57.7  &   20.76   &   0.94   &  3.1   & 2 &   0   &   0   &  -   &  -   &   1   &  \emph{1}   &   2   \\
   39611   &    03:27:19.33  -27:37:34.1  &   20.31   &   0.89   &  3.4   & 2 &   0   &   -   &  -   &  -   &   1   &  -   &   1   \\
   51155   &    03:27:19.35  -27:30:24.2  &   21.37   &   0.92   &  4.2   & 2 &   0   &   -   &  -   &  -   &   1   &  -   &   1   \\
   64257   &    03:29:41.45  -27:22:05.8  &   20.09   &   0.89   &  4.7   & 2 &   0   &   -   &  -   &  -   &   0   &  -   &   0   \\
   69697   &    03:30:47.23  -27:18:29.9  &   21.65   &   0.03   &  3.2   & 2 &   0   &   -   &  -   &  -   &   0   &  -   &   0   \\
   72745   &    03:30:41.20  -27:16:18.7  &   19.86   &   0.89   &  3.3   & 2 &   0   &   -   &  -   &  -   &   1   &  -   &   1   \\
   73492   &    03:28:47.12  -27:15:54.1  &   21.14   &   0.92   &  5.6   & 2 &   0   &   -   &  -   &  -   &   1   &  -   &   1   \\
   76729   &    03:28:42.85  -27:13:57.3  &   19.98   &   0.89   &  4.2   & 2 &   0   &   -   &  -   &  -   &   1   &  -   &   1   \\
   78591   &    03:30:51.47  -27:12:54.8  &   19.84   &   0.86   &  5.0   & 2 &   0   &   -   &  -   &  -   &   1   &  -   &   1   \\
   81775   &    03:28:45.45  -27:11:17.3  &   21.29   &   0.94   &  4.5   & 2 &   0   &   -   &  -   &  -   &   1   &  -   &   1   \\
   81788   &    03:27:23.50  -27:11:14.0  &   19.95   &   0.9   &  3.2   & 2 &   0   &   -   &  -   &  -   &   1   &  -   &   1   \\
   2663   &    03:31:35.44  -28:03:15.9  &   21.24   &   0.9   &  3.1   & 2 &   0   &   0   &  1   &  1   &   1   &  1   &   4   \\
   13037   &    03:35:37.95  -27:57:23.8  &   21.36   &   0.94   &  3.0   & 2 &   0   &   0   &  -   &  -   &   1   &  \emph{1}   &   2   \\
   19168   &    03:35:02.70  -27:53:33.4  &   20.28   &   0.03   &  3.2   & 2 &   0   &   0   &  -   &  -   &   1   &  \emph{0}   &   1   \\
   21995   &    03:31:35.77  -27:51:35.0  &   21.41   &   0.92   &  3.3   & 2 &   0   &   0   &  1   &  1   &   1   &  1   &   4   \\
   48715   &    03:32:59.19  -27:36:11.8  &   21.62   &   0.9   &  3.5   & 2 &   0   &   -   &  1   &  -   &   1   &  1   &   3   \\
   52751   &    03:34:45.88  -27:34:00.5  &   21.91   &   0.91   &  3.2   & 2 &   0   &   0   &  -   &  -   &   1   &  \emph{0}   &   1   \\
   76002   &    03:34:40.20  -27:21:38.1  &   20.84   &   0.04   &  8.0   & 2 &   0   &   -   &  -   &  -   &   1   &  -   &   1   \\
   76942   &    03:32:19.42  -27:21:03.0  &   20.95   &   0.85   &  3.5   & 2 &   0   &   -   &  -   &  -   &   1   &  -   &   1   \\
   78027   &    03:34:12.02  -27:20:29.9  &   19.9   &   0.08   &  3.3   & 2 &   0   &   -   &  -   &  -   &   1   &  -   &   1   \\
   79292   &    03:33:54.77  -27:19:42.8  &   21.08   &   0.96   &  4.7   & 2 &   0   &   -   &  -   &  -   &   1   &  -   &   1   \\
   86211   &    03:32:00.12  -27:15:46.2  &   19.84   &   0.03   &  3.3   & 2 &   0   &   -   &  -   &  -   &   0   &  -   &   0   \\
   90077   &    03:31:38.46  -27:13:51.8  &   20.17   &   0.03   &  4.5   & 2 &   1   &   -   &  -   &  -   &   0   &  -   &   -2   \\
   103927   &    03:32:14.61  -27:06:51.3  &   20.97   &   0.9   &  4.7   & 2 &   0   &   -   &  -   &  -   &   1   &  -   &   1   \\
   104734   &    03:32:01.24  -27:06:28.5  &   20.47   &   0.03   &  3.5   & 2 &   0   &   -   &  -   &  -   &   1   &  -   &   1   \\
   24635   &    03:29:03.71  -27:47:08.1  &   21.65   &   0.03   &  8.1   & 2 &   0   &   0   &  -   &  -   &   -   &  \emph{0}   &   0   \\

\end{longtable}
}% End onllongtab
}

\begin{acknowledgements}
      We acknowledge the Department of Physics of the University Federico II Naples, Italy. 
GL and EC acknowledge the financial support by PRIN 2011 'Cosmology with Euclid Mission'. We acknowledge PRIN INAF 2011 `Galaxy evolution with VST'; we also acknowledge the support from PRIN 2014 (P.I. N. Napolitano). 

This work was supported by the European Commission Research Executive Agency
FP7-SPACE-2013-1 Scheme (Grant Agreement 607254 - Herschel Extragalactic Legacy
Project - HELP) and by the Italian Ministry for Foreign Affairs and International Cooperation Joint Research Projects of Particular Relevance (Grant Agreement
ZA14GR02 - `Mapping the Universe on the Pathway to SKA').

The authors acknowledge L. Greggio for support at OmegaCAM. 

Funding for SDSS-III has been provided by the Alfred P. Sloan Foundation, the Participating Institutions, the National Science Foundation, and the U.S. Department of Energy Office of Science. The SDSS-III web site is http://www.sdss3.org/.
 SDSS-III is managed by the Astrophysical Research Consortium for the Participating Institutions of the SDSS-III Collaboration including the University of Arizona, the Brazilian Participation Group, Brookhaven National Laboratory, Carnegie Mellon University, University of Florida, the French Participation Group, the German Participation Group, Harvard University, the Instituto de Astrofisica de Canarias, the Michigan State/Notre Dame/JINA Participation Group, Johns Hopkins University, Lawrence Berkeley National Laboratory, Max Planck Institute for Astrophysics, Max Planck Institute for Extraterrestrial Physics, New Mexico State University, New York University, Ohio State University, Pennsylvania State University, University of Portsmouth, Princeton University, the Spanish Participation Group, University of Tokyo, University of Utah, Vanderbilt University, University of Virginia, University of Washington, and Yale University.
G.P. acknowledges support provided by the Millennium Institute of
Astrophysics (MAS) through grant IC120009 of the Programa Iniciativa
Cientifica Milenio del Ministerio de Economia, Fomento y Turismo de Chile.\\
We thank the anonymous referee for constructive comments that were helpful for improving the paper. 
\end{acknowledgements}

\bibliographystyle{aa}
\bibliography{bibtex}

\end{document}